\documentclass[twocolumn]{autart}    

\usepackage{graphicx}          
\usepackage[dvips]{epsfig}    
\usepackage{amsfonts}       
\usepackage{amssymb}
\usepackage{amsmath}
\usepackage{color}
\usepackage{wrapfig}
\newtheorem{thrm}{Theorem}
\newtheorem{lemm}{Lemma}
\newtheorem{deffn}{Definition}
\newtheorem{propo}{Proposition}
\newtheorem{assump}{Assumption}
\newtheorem{corol}{Corollary}

\newtheorem{remark}{Remark}
\newtheorem{example}{Example}

\def \det {\mathrm{det}}

\def \exp {\mathrm{exp}}

\def \var {\mathrm{Var}}
\def \vec {\mathrm{vec}}

\def \R {\mathbb{R}}
\def \C {\mathbb{C}}

\def \tailconst {b}
\def \tailcoeff {d}
\def \tailexp {\alpha}

\def \ssconstant {c}
\def \endproof {\hfill Q.E.D. $\blacksquare$}

\newcommand{\Mnorm}[2]{{\left\vert\kern-0.25ex\left\vert\kern-0.25ex\left\vert #1 
		\right\vert\kern-0.25ex\right\vert\kern-0.25ex\right\vert}_{#2}}
\newcommand{\Opnorm}[3]{{\left\vert\kern-0.25ex\left\vert\kern-0.25ex\left\vert #1 
		\right\vert\kern-0.25ex\right\vert\kern-0.25ex\right\vert}_{#2 \to #3}}
\newcommand{\norm}[2]{{\left\vert\kern-0.25ex\left\vert #1 
		\right\vert\kern-0.25ex\right\vert}_{#2}}
\newcommand{\eigmax}[1]{\left| \lambda_{\max} \left( #1 \right)\right|}
\newcommand{\eigmin}[1]{\left| \lambda_{\min} \left( #1 \right)\right|}
\newcommand{\tr}[1]{\mathrm{tr} \left( #1 \right)}
\newcommand{\PP}[1]{\mathbb{P} \left(#1\right)}
\newcommand{\E}[1]{\mathbb{E} \left[#1\right]}

\newcommand{\MJordanconst}[2]{\eta_{#1}\left(#2\right)}
\newcommand{\mult}[1]{\mu\left(#1\right)}
\newcommand{\samplesize}[3]{{N}_{#1}\left(#2,#3\right)}
\newcommand{\event}[1]{{\mathcal{#1}}}
\newcommand{\noisemax}[2]{{\nu}_{#1}\left(#2\right)}
\newcommand{\rank}[1]{\mathrm{rank}\left(#1\right)}
\newcommand{\diag}[1]{\mathrm{diag}\left(#1\right)}
\newcommand{\mincoor}[1]{\left[#1\right]_{\min}}
\newcommand{\minpoly}[1]{\phi \left(#1\right)}
\newcommand{\innerproductmin}[2]{\psi \left(#1,#2\right)}
\newcommand{\innerproductminconstant}[1]{\psi \left(#1\right)}
\newcommand{\zinfinitybound}[2]{\xi \left(#1,#2\right)}
\newcommand{\zinfinityconstant}[1]{\xi \left(#1\right)}

\newcommand{\loss}[3]{\mathcal{L}_{#1} \left(#3\right)}
\newcommand{\trans}[1]{A_{#1}}
\newcommand{\esttrans}[1]{\hat{A}^{(#1)}}

\newcommand{\symmetrizer}[1]{\Phi \left(#1\right)}
\newcommand{\predbound}[2]{{\beta}_{#1}\left(#2\right)}

\newcommand{\dimension}[2]{\mathrm{dim}_{#1} \left(#2\right)}
\newcommand{\lyap}[1]{\kappa \left(#1\right)}
\newcommand{\term}[1]{\mathbb{T}_{#1}}

\newcommand{\edit}[1]{\textcolor{black}{#1}}

\begin{document}

\begin{frontmatter}
	
	\title{Finite Time Identification in Unstable Linear Systems} 

	\author[]{Mohamad Kazem Shirani Faradonbeh},    
	\author[]{Ambuj Tewari},               
	\author[]{George Michailidis}  

	\begin{keyword}                           
		Unstable Systems; Linear Dynamics; Finite Time Identification; Stabilization; Autoregressive Process; Non-Asymptotic Estimation               
	\end{keyword}                             
	
\begin{abstract}
	Identification of the parameters of stable linear dynamical systems is a well-studied problem in the literature, both in the low and high-dimensional settings. However, there are hardly any results for the unstable case, especially regarding {\em finite time bounds}. For this setting, 
	classical results on least-squares estimation of the dynamics parameters are not applicable and therefore new concepts and technical approaches
	need to be developed to address the issue. 
	Unstable linear systems arise in key real applications in control theory, econometrics, and finance. 
	
	This study establishes finite time bounds for the identification error of the least-squares estimates for a fairly large class of heavy-tailed noise distributions, and transition matrices of such systems.
	The results relate the time length (samples) required for estimation to a function of the problem dimension and key characteristics of the true underlying transition matrix and the noise distribution. To establish them, appropriate concentration inequalities for random matrices and for sequences of martingale differences are leveraged.
\end{abstract}
\end{frontmatter}


\section{Introduction}
Identification of the transition matrix in linear dynamical systems has been extensively studied in the literature for the stable case \cite{lutkepohl2005new,ljung1999system,soderstrom1989system}. Further, new work has also addressed this topic under a  high-dimensional scaling, with additional assumptions on sparsity of the parameters imposed on it \cite{basu2015regularized,zorzi2016ar,zorzi2017sparse}. However, in settings where the underlying dynamics are {\em not stable}, this problem 
has {\em not} been adequately studied. 
A key issue that arises in this case is that the magnitude of the state vector explodes with high probability, exponentially over time \cite{lai1985asymptotic}. Nevertheless, identification of the dynamics in the non-stable case is of interest due to a number of applications that give rise to such dynamics.  In addition to adaptive control \cite{soderstrom2012discrete,kailath2000linear,kumar2015stochastic,bertsekas1995dynamic}, these applications include a class of identification problems involving asset bubbles and high inflation episodes \cite{pesaran2005small,pesaran2010predictability,pesaran2002market,alogoskoufis1991phillips,garcia1991analysis,stock1996evidence,stock1998comparison,giacomini2006tests,nielsen2008explosive,engsted2006explosive,juselius2002high,nielsen2010analysis}. 

\edit{Most existing work on the topic provides {\em asymptotic} results on the convergence \cite{lai1985asymptotic}, as well as the limit distribution \cite{buchmann2007asymptotic,buchmann2013unified} of the model parameters.} Specifically,
early work investigated the limit distribution of the state vector under a set of restrictive assumptions on the dynamics matrix \cite{anderson1959asymptotic}. Ensuing work dealt
with the accuracy of identification in infinite time, for a class of structured transition matrices \cite{lai1983asymptotic}. Further extensions to more general
classes were established by Nielsen \cite{nielsen2005strong,nielsen2006order}. Finally, additional asymptotic results together with the important concept of {\em irregularity} of the  transition matrix which leads to inconsistency, are presented in the literature
\cite{nielsen2009singular}. However, finite time (i.e. non-asymptotic) results are not currently available.

In this work, we consider a linear dynamical system $x(t) \in \R^p, t=0,1,\cdots$ that evolves according to the following Vector Autoregressive (VAR) model
\begin{eqnarray} \label{VARdynamics}
	x(t+1)=\trans{0} x(t)+w(t+1), 
\end{eqnarray}
starting from an arbitrary initial state $x(0)$, \edit{which can be either deterministic or stochastic}. Note that systems of longer but finite memory can also be written in the above form \cite{soderstrom2012discrete,kailath2000linear}. We examine the general case where the system is not necessarily stable. The key contributions are: (i) establishing finite time identification bounds for the $\ell_2$ error of the least-squares estimates of the
transition matrix $\trans{0}$, (ii) under a fairly general heavy tailed noise (disturbance) process $\left\{w(t)\right\}_{t=1}^\infty$. In addition, the results due to the presence of a heavy-tailed 
noise term are of independent interest for the stable case as well. The novel results established provide insights
on how the time length required for identification scales both with the dimension of the system, as well as with the characteristics of the transition matrix and the noise process.

In order to establish results for accurate finite time identification of $\trans{0}$, one needs to address the following set of technical issues. Note that as long as $\trans{0}$ has eigenvalues outside of the unit circle in the complex plane, the behavior of the Gram matrix of the state vector is governed by a random matrix. However, when $\trans{0}$ has eigenvalues both inside and outside of the unit circle, the smallest eigenvalue of the Gram matrix scales linearly over time, while its largest eigenvalue grows exponentially, which in turn leads to the failure of the classical approaches to establish accurate identification. These issues are addressed in Subsections \ref{explosivecase} and \ref{generalcase}, respectively. In the proofs, we leverage selected concentration inequalities for random matrices \cite{tropp2012user}, as well as an anti-concentration property of martingale difference sequences \cite{lai1983note}.

The problem of fast accurate identification in unstable systems has a number of interesting applications. For example, in stochastic control, this includes the canonical problems of both stabilization, as well as design of an efficient adaptive policy for linear systems. First, since the dynamics are governed by unknown transition matrices, the control action can destabilize the system. Moreover, the user first needs to have an approximation of the dynamics, to be able to design a suitable control policy. Therefore, {\em accurate} identification of the dynamics of the transition matrices is necessary, even if they happen to lead to instability of the underlying system. More importantly, the identification result needs to be provided within a relative {\em short} time period for the user to be able to design the adaptive policy accordingly. More details are discussed in Example \ref{adaptiveexample}.

\edit{Applications of this setting in econometrics and finance also create the need to obtain finite time theoretic results. For example, in macroeconomics, the outstanding performance of the linear models marked them as a benchmark of forecasting the market \cite{pesaran2005small,stock1998comparison,giacomini2006tests}. Their applications to the analysis of inflationary episodes in a number of OECD\footnote{Organization for Economic Co-operation and Development} countries \cite{pesaran2005small}, as well as US stock prices \cite{engsted2006explosive,lin2017regularized} are available in the literature. The former study establishes the structural non-stationarity of the process, where the latter verifies the explosive behavior of speculative bubbles. In particular, if a technology market is capable of important innovations with uncertain outcomes, it has been argued \cite{pesaran2010predictability} that a bubble is very likely to emerge.}

\edit{Another application involving unstable dynamics deals with episodes of hyperinflation. For example, Juselius and Mladenovic \cite{juselius2002high} consider the case of (former) Yugoslavia and use data on various economic indicators to gain insights into 
the dynamics of the late 1990s episode. The analysis identifies wages, price level expectations, and currency depreciation as the key factors. In follow-up work, infinite time analysis techniques were used \cite{nielsen2010analysis}, but as emphasized in the original work \cite{juselius2002high} ``hyperinflation episodes almost by definition are {\em short}." Therefore, the small sample size available can easily lead to incorrect inference, while finite time guarantees are informative
about the sample size needed to make precise statements about the effects of different macroeconomic factors. Another hyperinflation episode from Germany in the early 1920's is studied by Nielsen \cite{nielsen2008explosive}.}

Recently, the problem of forecasting non-stationary mixing \cite{kuznetsov2017generalization,kuznetsov2014generalization}, and non-mixing \cite{kuznetsov2015learning} time series has received attention, assuming the loss function employed is bounded. Unstable VAR models are a special, yet interesting, case of non-stationary time series. However, the problem of estimation/identification is not still addressed in the existing literature. 
Moreover, the results on forecasting are not applicable to the identification problem, since the least-squares loss function used in that study is not bounded.  On the other hand, the obtained results on identification are applicable to forecasting.

The remainder of the paper is organized as follows. In Section \ref{Autoregressive Processes} we provide a rigorous formulation of the problem, introduce the identification procedure, and outline examples that require accurate identification but the system can not assumed to be stable. The contributions are discussed in Section \ref{mainresults}, where we study different scenarios. First, we provide identification results on (non-stationary) stable linear systems in Subsection \ref{stablecase}, followed by the explosive case (Subsection \ref{explosivecase}). Finally, we study the accurate identification of the dynamics for general systems in Subsection \ref{generalcase}.

\subsection{Notations}
The following {notation} is used throughout this paper. For a matrix $A \in \C^{p \times q}$, $A'$ denotes its transpose. When $p=q$, the smallest (respectively largest) eigenvalue of $A$ (in magnitude) is denoted by $\lambda_{\min} (A)$ (respectively $\lambda_{\max}(A)$). For $\gamma \in \R, \gamma > 0, x \in \C^q$, define the norm $\norm{x}{\gamma} = \left( \sum\limits_{i=1}^{q} \left| x_i \right|^\gamma \right)^{1/\gamma}$. For $\gamma = \infty$, define the norm $\norm{x}{\infty} = \max \limits_{1 \leq i \leq q} |x_i|$. 

\edit{We also use the following notation for the operator norm of matrices. For $\beta, \gamma \in \left(0,\infty\right], A \in \C^{p \times q}$ let, 
\begin{equation*}
\Opnorm{A}{\gamma}{\beta} = \sup \limits_{v \in \C^{q} \setminus \{0\}} \frac{\norm{Av}{\beta}}{\norm{v}{\gamma}}.
\end{equation*}
Whenever $\gamma = \beta$, we simply write $\Mnorm{A}{\beta}$. To show the dimension of manifold $M$ over the field $F$, we use $\dimension{F}{M}$. The sigma-field generated by random vectors $X_1,\cdots,X_n$ is denoted by $\sigma \left( X_1,\cdots,X_n \right)$. Finally, the symbol $\vee$ denotes the maximum of two or more quantities.}

\section{Problem Formulation and Preliminaries} \label{Autoregressive Processes}
The system $\left\{x(t)\right\}_{t=0}^\infty$ evolves according to \eqref{VARdynamics}, while the unknown transition matrix $\trans{0} \in \mathbb{R}^{p \times p}$ is not assumed to be stable, i.e. the eigenvalues of $\trans{0}$ do not necessarily lie inside the unit circle. Further, $\left\{w(t)\right\}_{t=1}^\infty$ is the sequence of independent mean-zero noise vectors with covariance matrix $C$, i.e. $\E {w(t)}=0$, and $\E {w(t)w(t)'}=C$. 
\begin{remark}
	The results established also hold if the noise vectors are martingale difference sequences. Further, the generalization to heteroscedastic noise, where the covariance matrix $C$ is time varying, is rather straightforward.
\end{remark}

\edit{The objective is to identify $\trans{0}$, using the least-squares estimator. One observes the state vector during a finite time interval, $\left\{x(t)\right\}_{t=0}^n$, and defines the sum-of-squares loss function
	\begin{eqnarray*}
		\loss{n}{}{\trans{}}=\sum\limits_{t=0}^{n-1} \norm{x(t+1)-\trans{}x(t)}{2}^2.
	\end{eqnarray*}
	Then, $\trans{0}$ is estimated by $\esttrans{n}$, which is the minimizer of the above sum-of-squares; $\loss{n}{}{\esttrans{n}} = \min\limits_{\trans{} \in \R^{p \times p }} \loss{n}{}{\trans{}}$.}

The {main} contribution of this paper is to establish that with high probability, accurate identification of the true transition matrix is achieved, excluding a pathological case. Formally, \edit{for arbitrary accuracy $\epsilon>0$ and failure probability $\delta>0$,} $\esttrans{n}$ is with probability at least $1-\delta$ within an $\epsilon$-neighborhood of $\trans{0}$, where apart from a logarithmic factor, the time length $n$ scales quadratically with $\epsilon^{-1}$, and logarithmically with $\delta^{-1}$. \edit{In other words, for a fixed accuracy $\epsilon>0$, the probability that the identification error $\Mnorm{\esttrans{n}-\trans{0}}{2}$ exceeds $\epsilon$, decays {\em exponentially} as $n$ grows.}

The following example elaborates on the problem of finite time identification for unstable dynamical systems in control theory. 
\begin{example}[Stabilization in adaptive control] \label{adaptiveexample}
	Consider the linear stochastic system $\left[ \trans{x}, \trans{u}\right]$, where the state evolution is governed by the following dynamics:
	\begin{eqnarray*}
		x(t+1)=\trans{x}x(t)+\trans{u}u(t)+w(t+1).
	\end{eqnarray*}
\end{example}

In the previous equation, the vector $x(t) \in \R^p$ represents the state of the system, and $u(t) \in \R^r$ is the control action taken by the user. The {\em unknown} transition matrix $\trans{x} \in \R^{p \times p}$ determines the evolution of the system, and the {\em unknown} input matrix $\trans{u} \in \R^{p \times r}$ shows the effect of the control policy on the state of the system. 

Due to the simplicity of the structure, the main interest is in linear feedbacks of the form $u(t)=Lx(t)$, where $L \in \R^{r \times p}$ is the feedback matrix. Further, in addition to preserving the linear nature of the system (which prevents the analysis from becoming mathematically intractable), linear feedbacks correspond to important objectives for a class of optimal control problems \cite{soderstrom2012discrete,kumar2015stochastic}, including minimization of quadratic costs \cite{bertsekas1995dynamic,brunner2018stochastic}. So, the linear dynamics are essentially determined by the closed-loop transition matrix $\trans{0}=\trans{x}+\trans{u}L$.

\begin{figure}[t!] 
	\centering
	\scalebox{.5}
	{\includegraphics {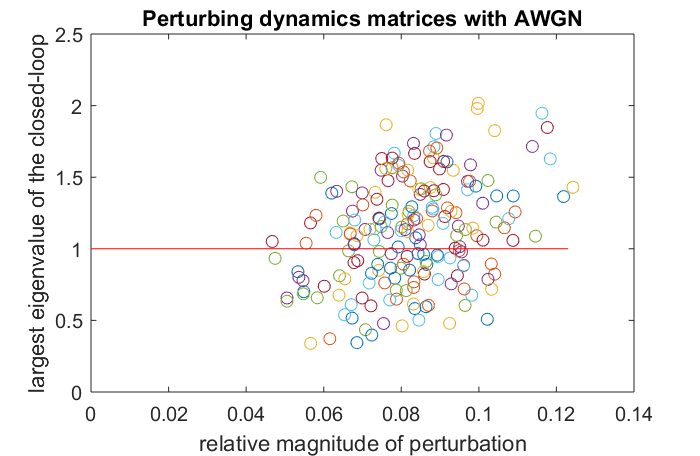}} 
	\caption{\edit{$\eigmax{A_0}$ vs $\norm{\Delta}{2}/\norm{\Theta_0}{2}$.}}
	\label{Fig1}
\end{figure}

\edit{The system $\Theta_0=\left[ \trans{x}, \trans{u} \right]$ is assumed to be stabilizable, implying there exists a stabilizer $L_0$ such that the closed-loop matrix $\trans{x}+\trans{u}L_0$ is stable; $\eigmax{\trans{x}+\trans{u}L_0}<1$. Finding such a stabilizer requires precise approximation of the true dynamics $\Theta_0$ \cite{bertsekas1995dynamic}, as shown in the following example. Consider a system of dimension $p=3, r=2$, 
which is stabilizable, since exact knowledge of $\Theta_0$ yields $\eigmax{\trans{0}}= 0.22$. Fig. \ref{Fig1} depicts the scatter plot of the largest eigenvalue of the closed-loop transition matrix versus the relative magnitude of an Additive White Gaussian Noise (AWGN). A stabilizing linear feedback $L$ is applied to the system as if the dynamics parameter is $\Theta_0 + \Delta $ instead of $\Theta_0$, where entries of $\Delta$ are independent Gaussian measurement errors. It can be seen that a measurement error as small as $5\%$ in the identification of the system dynamics can lead to instability.}

\begin{figure}[t!] 
	\centering
	\scalebox{.5}
	{\includegraphics {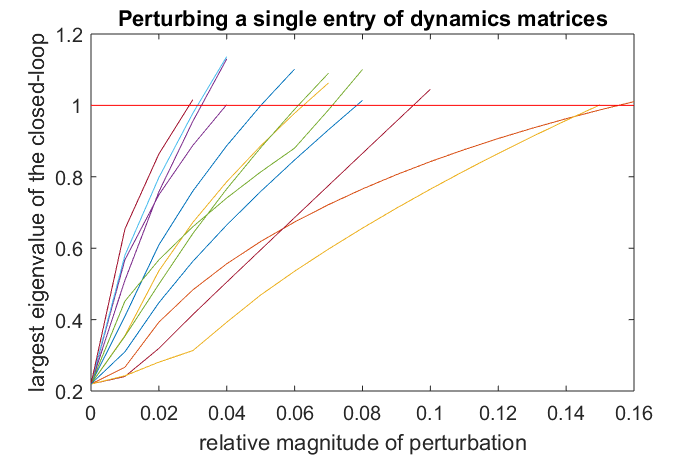}} 
	\caption{\edit{$\eigmax{A_0}$ vs $\epsilon/\norm{\Theta_0}{2}$. }}
	\label{Fig2}
\end{figure}

\edit{Fig. \ref{Fig2} graphs the largest eigenvalue of the closed-loop matrix versus a perturbation in a single entry of $\Theta_0$. In fact, for different entries of $\Theta_0$, the linear feedback $L$ is designed as if the operator approximates a single entry incorrectly. Formally, for $\epsilon \geq 0$, only the $(i,j)$-th entry of $\Theta_0$ is approximated with error $\epsilon$, while all other entries are exactly provided to the operator. Fig. \ref{Fig2} corresponds to the relationship between $\eigmax{\trans{0}}$ and $\epsilon$, for different entries $(i,j)$. Therefore, stabilization is very sensitive to the perturbation, as an error of $3\%$ in relative magnitude in a single element of the system will totally destabilize the system. In many applications, especially if the system under consideration is not man-made, such precise information is not available. Hence, the matrix $\trans{0}$ can not be assumed to be a priori stable.}

In addition, in order to design a desired policy (either steering the system to a specific state \cite{faradonbeh2016optimality} or minimizing a cost function \cite{bertsekas1995dynamic}), such an approximation is necessary. To obtain it, learning accurately the dynamics of an unstable system is needed. Importantly, such learning needs to conclude in  finite time, because afterwards, the user needs to control the system to achieve the corresponding objective, determined by the application. 

In order to establish high probability guarantees for accurate identification of the closed-loop matrix, we apply the results from Theorem \ref{consistency} given in the next Section. A random linear feedback, denoted by $L$, suffices to satisfy the assumptions of Theorem \ref{consistency} for the closed-loop matrix $\trans{x}+ \trans{u} L$. In fact, it suffices for $L$ to be a continuously distributed random matrix. This in turn, leads to accurate identification of $\left[ \trans{x}, \trans{u} \right]$, applying multiple random linear feedbacks, drawn independently. Note that direct identification of $\left[ \trans{x}, \trans{u} \right]$ is infeasible, since by observing the state sequence $\left\{ x(t) \right\}_{t=0}^\infty$, the best result one can provide is ``closed-loop identification" \cite{soderstrom2012discrete,kumar1990convergence}. \edit{Specifically, for a given closed-loop transition matrix $\trans{0}$, the set of parameters guiding the system's dynamics $\trans{x},\trans{u}$ which satisfy $\trans{0}=\trans{x}+\trans{u}L$ is not unique if one exactly knows the feedback matrix $L$. This set is indeed a subspace of dimension $pr$ in the space $\R^{p \times (p+r)}$ the matrix $\left[\trans{x},\trans{u}\right]$ belongs to.}

To analyze the finite time behavior of the aforementioned identification procedure, the following is assumed for the tail-behavior of every coordinate of the noise vector.
\begin{assump}[Sub-Weibull noise distribution] \label{tailcondition}
	There exist positive constants $\tailconst, \tailcoeff$, and $\tailexp$, such that for all $t=1,2,\cdots; i=1, \cdots, p; y >0$, 
	\begin{eqnarray*}
		\PP{\left|w_i(t)\right| > y} \leq \tailconst \: \exp \left(-\frac{y^\tailexp}{\tailcoeff}\right).
	\end{eqnarray*}
\end{assump}

\edit{In case of random initial state $x(0)$, we assume it also follows a sub-Weibull distribution.} Intuitively, smaller values of the exponent $\tailexp$ correspond to heavier tails for the noise distribution, and vice versa. Note that assuming a sub-Weibull distribution for the noise coordinates is more general than the sub-Gaussian (or sub-exponential) assumption routinely made in the literature \cite{abbasi2011online}, where $\tailexp \geq 2$ ($\tailexp \geq 1$). In fact, when $\tailexp <1$, the noise coordinates $w_i(t)$ do not need to have a moment generating function. 

\edit{Note that for establishing consistency of infinite time identification procedures, the noise vectors need to satisfy a moment condition, e.g. $\E{\norm{w(t)}{2}^{2+\tailexp}}<\infty$, for some $\tailexp>0$ \cite{lai1985asymptotic,lai1983asymptotic}. On the other hand, finite time identification analysis results are usually obtained under an assumption of a light-tail (or even uniformly bounded) noise distribution; e.g. Gaussian process \cite{tropp2012user,abbasi2011online}. Thus, the above assumption on sub-Weibull noise, that includes a family of heavy-tailed noise processes, provides a fairly general framework to narrow down the theoretical gap between asymptotic and non-asymptotic approaches. }

The noise coordinates can be either discrete or continuous random variables, and are not assumed to have a probability density function. To proceed, we define a property of the population covariance matrix of the system under study. It is easy to see that the following property is necessary and sufficient for accurate estimation of dynamics parameters. 

\begin{deffn}[Reachability] \label{reachability}
	The pair $\left[\trans{0},C\right]$ is called reachable if 
	\begin{eqnarray*}
		\rank{\left[C^{1/2}, \trans{0}C^{1/2}, \cdots, \trans{0}^{p-1}C^{1/2} \right]}=p.
	\end{eqnarray*} 
\end{deffn}
Clearly, reachability is equivalent to $\eigmin{K(C)}>0 $, where $K(C)=\sum\limits_{i=0}^{p-1}\trans{0}^i C {\trans{0}'}^i$.
Specifically, if $C$ is positive definite, then $\left[\trans{0},C\right]$ is reachable for all $\trans{0} \in \R^{p\times p}$.
Reachability is conceptually equivalent to the population covariance matrix of the system being positive definite. More precisely, since the noise vectors are independent, the covariance matrix of $x(t)$ is given by $\sum\limits_{i=0}^{t-1}\trans{0}^i C {\trans{0}'}^i$; i.e. reachability is in fact stating that for $t \geq p$, every coordinate of $x(t)$ has non-degenerate randomness. 

\edit{Further, reachability is particularly helpful if the actual evolution of the system is guided by VAR$\left( k \right)$ dynamics, for some $k>1$. In this case, the next step is determined by the $k$ previous lags: for $t \geq k$, the state sequence $\tilde{x}(t) \in \R^m$ evolves according to
	\begin{eqnarray*}
		\tilde{x}(t) = \sum\limits_{j=1}^{k} \trans{j} \tilde{x}(t-j) + \tilde{w}(t),
	\end{eqnarray*}
	for some initial vectors $\tilde{x}(0), \cdots, \tilde{x}(k-1) \in \R^m$, and transition matrices $\trans{1}, \cdots, \trans{k} \in \R^{m \times m}$, assuming $\trans{k} \neq 0$. Arranging blocks of $\tilde{x}(t)$ accordingly, $x(t) = \left[ \tilde{x}(t+k-1)' ,  \cdots , \tilde{x}(t)' \right] ' \in \R^{km}$, the state evolution can be written in the form of \eqref{VARdynamics}, for $\trans{0}=\begin{bmatrix} \trans{1} \cdots \trans{k-1} & \trans{k} \\ I_{(k-1)m} & 0 \end{bmatrix} \in \R^{km \times km}$. Then, as long as the covariance matrix of $\tilde{w}(t)$ is full rank, reachability holds.}


\section{Main results} \label{mainresults}

Next, we establish the key identification results that characterize the time (samples) required, so that with high probability the $\trans{0}$ least-squares estimate is accurate within a certain degree. First, we study the identification for stable systems where all eigenvalues of $\trans{0}$ are inside the unit circle, i.e. $\eigmax{\trans{0}}<1$. Subsequently, the explosive case where all eigenvalues of the transition matrix $\trans{0}$ lie outside of the unit circle, i.e. $\eigmin{\trans{0}}>1$, is examined. Finally, finite time identification results are presented for the general case which is the combination of these two regimes.

Some straightforward algebra shows that the least-squares estimator can be written as
\begin{eqnarray*}
	\esttrans{n}= \sum\limits_{t=0}^{n-1} x(t+1)x(t)'V_n^{-1},
\end{eqnarray*}
where
$V_n=\sum\limits_{t=0}^{n-1} x(t)x(t)'$ denotes the empirical covariance matrix of the state process \edit{(once normalized by $n$)}, which is assumed to be non-singular.

The latter result implies that the behavior of $V_n$ needs to be carefully studied and this constitutes a major part of the following two subsections. 

\subsection{Stable systems} \label{stablecase}
The stable case has been extensively studied before, customarily under the stronger assumption of sub-Gaussian noise \cite{abbasi2011online}. Next, we generalize the results to sub-Weibull noise vectors defined in Assumption \ref{tailcondition}. Further, these results will be used for the general case in Subsection \ref{generalcase}.

\edit{For a stable transition matrix $\trans{0} \in \R^{p \times p}$, we define the constant $\MJordanconst{}{\trans{0}}$, that is critical in specifying various constants that appear in the main results.
Its definition is based on the Jordan decomposition of square matrices.} 
	
\edit{First, for $\lambda \in \C$, define the size $m$ Jordan matrix of $\lambda$ as follows.
	\begin{eqnarray*} \label{Jordanform}
		\begin{bmatrix}
			\lambda & 1 & 0 & \cdots & 0 & 0 \\
			0 & \lambda & 1 & 0 & \cdots & 0 \\
			\vdots & \vdots & \vdots & \vdots & \vdots & \vdots \\
			0 & 0 & \cdots & 0 & \lambda & 1 \\
			0 & 0 & 0 & \cdots & 0 & \lambda
		\end{bmatrix} \in \C^{m \times m}.
	\end{eqnarray*}
Then, the Jordan decomposition of $\trans{0}$ is given by 
$\trans{0}=P^{-1}\Lambda P$, where $\Lambda$ is block diagonal, $\Lambda= \diag{\Lambda_1,\cdots, \Lambda_k}$, with $\Lambda_i \in \mathbb{C}^{m_i \times m_i}, i=1, \cdots, k$ being a Jordan matrix of $\lambda_i$.
	\begin{deffn} \label{Jordandeff}
		For a stable matrix $\trans{0}$, suppose that $\trans{0}=P^{-1} \Lambda P$ is the Jordan decomposition as described above. For $t=1,2, \cdots$, let
		\begin{eqnarray*}
			\MJordanconst{t}{\Lambda_i}=\inf\limits_{\rho \geq \left|\lambda_i\right|} t^{m_i-1} \rho^t \sum\limits_{j=0}^{m_i-1} \frac{\rho^{-j}}{j!},
		\end{eqnarray*}
		and $\MJordanconst{t}{\Lambda}= \max \limits_{1 \leq i \leq k} \MJordanconst{t}{\Lambda_i}$. Then, letting $\MJordanconst{0}{\Lambda}=1$, define
		\begin{eqnarray*}
			\MJordanconst{}{\trans{0}} = \Opnorm{P^{-1}}{\infty}{2} \Mnorm{P}{\infty} \sum\limits_{t=0}^{\infty} \MJordanconst{t}{\Lambda}.
		\end{eqnarray*}
\end{deffn}
Clearly, denoting the largest algebraic multiplicity of the eigenvalues of $\trans{0}$ (which is the same to the largest block-size in the Jordan form) by $\mult{\trans{0}}=\max\limits_{1 \leq i \leq k} m_i$, we have
\begin{eqnarray} \label{Jordaneq1}
\MJordanconst{t}{\Lambda} \leq t^{\mult{\trans{0}}-1} \eigmax{\trans{0}}^t e^{\eigmax{\trans{0}}^{-1}}.
\end{eqnarray}
}

In the stable regime, the state process has a stationary limit distribution. In this case, the empirical covariance matrix has an approximately deterministic behavior, which is described by its asymptotic distribution. Specifically, as time grows, $V_n$ appropriately normalized, can be approximated by $\lyap{C}$, where $\lyap{C}= \sum\limits_{i=0}^{\infty} \trans{0}^i C {\trans{0}'}^i$ denotes the asymptotic covariance matrix. 

\edit{The following lemma provides a finite lower bound for the time length (number of samples), based on the identification error $\epsilon$, and the failure
probability $\delta$. For this purpose, using the parameters $\tailconst, \tailcoeff$, and $\tailexp$ specified in Assumption \ref{tailcondition}, we define the following constant. 
Henceforth, one can let $\tailexp \to \infty$, if the noise vectors $w(1),w(2), \cdots$ are uniformly bounded.
	\begin{eqnarray*}
		\ssconstant_1 &=& 288 \left( \norm{x(0)}{\infty} \vee 1 \right)^2 \MJordanconst{}{\trans{0}}^2 \MJordanconst{}{\trans{0}'}^4 \left( \Mnorm{\trans{0}}{2}^2 +1\right) \\
		&\times& \left( \eigmax{C}+1 \right) \left( \tailcoeff \log 2\tailconst p \right)^{4/\tailexp} p \log 8p.
	\end{eqnarray*}} 

\begin{lemm} \label{stablemin}
	Assuming $\eigmax{\trans{0}}<1$, let $\ssconstant_1 $ be as defined above. Then, for arbitrary $\epsilon, \delta>0$ if 
	\begin{eqnarray*} \label{stablemineq0}
		\frac{n}{\left(\log n\right)^{4/\tailexp}} \geq \frac{\ssconstant_1}{\epsilon^2} \left(-\log \delta\right)^{1+4/\tailexp},
	\end{eqnarray*} 
	then
	\begin{eqnarray*}
		\PP{\eigmax{\frac{1}{n}V_{n+1}- \lyap{C}} > \epsilon} \leq \delta.
	\end{eqnarray*}
\end{lemm}

A direct consequence of Lemma \ref{stablemin} is the following corollary, which shows that high probability accurate identification can be ensured, if reachability, as defined in Definition \ref{reachability}, is assumed. Note that reachability implies that $\lyap{C}$ is positive definite. \edit{Using $\ssconstant_1$ defined above, we define $\ssconstant_2= 4 \ssconstant_1 \left( \Mnorm{\trans{0}}{2}^2 \vee 1 \right) \eigmin{K(C)}^{-2}+2$.}
\begin{corol} \label{stableestimation}
	Suppose that $\eigmax{\trans{0}}<1$, and $\left[\trans{0},C\right]$ is reachable. Then, for $\ssconstant_2$ above, and for all $\epsilon,\delta>0$,  
	\begin{eqnarray*} 
		\frac{n}{\left(\log n\right)^{4/\tailexp}} \geq \frac{\ssconstant_2}{\epsilon^2} \left(-\log \delta\right)^{1+4/\tailexp},
	\end{eqnarray*} 
	implies 
	\begin{eqnarray*}
		\PP{\Mnorm{\esttrans{n}-\trans{0}}{2}>\epsilon} < \delta.
	\end{eqnarray*} 
\end{corol}

\subsection{Explosive systems} \label{explosivecase}
In the explosive case, the empirical covariance matrix $V_n$ grows exponentially with respect to $n$. In addition, unlike the stable case, $V_n$ appropriately normalized, can be approximated by a random matrix. Therefore, the eigenvalues of the normalized empirical covariance matrix are stochastic as well. In order to find deterministic bounds for the eigenvalues of $V_n$, new quantities, denoted by $\minpoly{\trans{0}}, \innerproductmin{\trans{0}}{\delta}$, need to be defined. 

Subsequently, after providing formal definitions of these quantities, we present in Lemma \ref{explosivemin} bounds for the eigenvalues. Then, a sufficient and necessary property of $\trans{0}$ for accurate identification is introduced, followed by Propositions \ref{minpoly}, \ref{PDcore}, which establish the positiveness of $\minpoly{\cdot}, \innerproductmin{\cdot}{\cdot}$. This subsection concludes with Corollary \ref{explosiveestimation} that deals with identification in explosive systems.

First, for explosive $\trans{0}$, we define the nonnegative functions $\minpoly{\trans{0}}, \innerproductmin{\trans{0}}{\delta}$ as follows. Assuming $\eigmin{\trans{0}}>1$, let $\trans{0}=P^{-1}\Lambda P$ be the Jordan decomposition. Letting 
\begin{eqnarray*}
	z(\infty) &=& x(0)+\sum\limits_{i=1}^{\infty}\trans{0}^{-i}w(i), \\
	P &=& \left[P_1, \cdots, P_p\right]',
\end{eqnarray*}
for $\delta>0$ define 
\begin{eqnarray*}
	\innerproductmin{\trans{0}}{\delta} = \sup \left\{ y \in \R : \PP{ \min\limits_{1 \leq i \leq p} \left|P_i'z(\infty)\right| < y} \leq \delta \right\}.
\end{eqnarray*}
Note that according to this definition, all coordinates of the vector $Pz(\infty)$ are in magnitude at least $\innerproductmin{\trans{0}}{\delta}$, with probability at least $1-\delta$. Next, define
\begin{eqnarray*}
	\minpoly{\trans{0}} = \Opnorm{P}{2}{\infty}^{-1} \inf \limits_{a \in {\R}^p \setminus \{0\}} \frac{1}{\norm{a}{1}} \mincoor{\sum\limits_{i=0}^{p-1}a_{i+1}\Lambda^{-i}},
\end{eqnarray*}
where for an arbitrary matrix $M \in \C^{m \times k}$, $\mincoor{M}$ is the smallest magnitude of the nonzero entries of $M$:
\begin{eqnarray*}
	\mincoor{M}= \min \{\left|M_{ij}\right|: 1 \leq i \leq m; \!\ 1 \leq j \leq k ; \!\ M_{ij} \neq 0 \}.
\end{eqnarray*}

In fact, $\minpoly{\trans{0}}$ represents the deterministic portion of the smallest eigenvalue of the random matrix $F_\infty$ which approximates the normalized $V_n$. It only depends on $\trans{0}$, while $\innerproductmin{\trans{0}}{\delta}$ represents the stochastic portion which depends on both $\trans{0}$ and the distribution of the noise sequence $\left\{w(t)\right\}_{t=1}^\infty$. Intuitively, $\minpoly{\trans{0}}$ denotes the minimum nontrivial distance between the polynomials of $\trans{0}^{-1}$ and the origin, and $\innerproductmin{\trans{0}}{\delta}$ denotes the high probability minimum distance of the vector $Pz(\infty)$ from the origin. These minimum distances show up, because for $v \in \R^p$, $v'F_\infty v$ is determined by the product of a polynomial of $\trans{0}^{-1}$ (with coefficients determined by $v$), and $Pz(\infty)$. More details are provided in the proof of Lemma \ref{explosivemin}.

Now, the behavior of the normalized empirical covariance matrix can be controlled as follows:
\begin{lemm} \label{explosivemin}
	Suppose that $\eigmin{\trans{0}}>1$; then, there is a constant $\zinfinityconstant{\trans{0}} < \infty$ such that for all $n,\delta$,
	\begin{eqnarray*}
		\PP{\eigmax{\trans{0}^{-n} V_{n+1} {\trans{0}'}^{-n}} > \zinfinityconstant{\trans{0}} \left(-\log\delta\right)^{2/\tailexp}} \leq \delta.
	\end{eqnarray*}
	\edit{Further, there is a constant $n_1 < \infty$, such that for arbitrary $\epsilon,\delta>0$ if
		\begin{eqnarray} \label{SSexplosivemin}
		n \geq \frac{3 \left( \tailexp+2 \right)}{\tailexp \log \eigmin{\trans{0}}} \log \left(\frac{- \log \delta}{\epsilon} \right) \vee n_1,
		\end{eqnarray}
		then with probability at least $1-4\delta$ it holds that
		\begin{eqnarray} \label{explosivemineq1}
		\eigmin{\trans{0}^{-n} V_{n+1} {\trans{0}'}^{-n}} \geq \minpoly{\trans{0}}^2\innerproductmin{\trans{0}}{\delta}^2 - \epsilon.
		\end{eqnarray}}
\end{lemm}

\begin{remark} \label{explosiveminremark}
	The inequality \eqref{SSexplosivemin} is of interest for the following two reasons. First, the accuracy $\epsilon$ decays exponentially fast when $n$ grows. Second, the failure probability $\delta$ decays {\em doubly} exponentially fast with $n$. 
\end{remark}
This surprising strong behavior is intuitively caused by the exponential growth of $x(t)$. Broadly speaking, the growing signal (i.e. $x(t)$) to noise (i.e. $w(t)$) ratio leads to the super fast decay of $\epsilon$ and $\delta$. Note that commonly in identification problems, the decay rates of $\epsilon, \delta$ are square-root, and exponential, respectively.

If $\minpoly{\trans{0}} \innerproductmin{\trans{0}}{\delta}=0$, obviously \eqref{explosivemineq1} holds. Thus, the main interest is in the case where $\minpoly{\trans{0}} \innerproductmin{\trans{0}}{\delta} \neq 0$, which we will show that holds under certain conditions, and is necessary to ensure accurate identification. In fact, the first case is of no interest, since it can be shown that $V_n$ will be singular, and thus identification of $\trans{0}$ fails, even if the time period becomes infinitely large \cite{nielsen2009singular}. For the second case, the transition matrix $\trans{0}$ needs to be regular, according to the following definition. Regularity (of course in addition to reachability), leads to accurate identification as
shown in Corollary \ref{explosiveestimation}. 

\begin{deffn}[Regularity] \label{regularity}
	$\trans{} \in \R^{p \times p}$ is called regular if for any explosive eigenvalue of $\trans{}$, denoted by $\lambda$, the geometric multiplicity of $\lambda$ is one.
\end{deffn}

Regularity essentially implies that the eigenspace corresponding to $\lambda$ is one dimensional. There are also equivalent formulations for regularity. Indeed, $\trans{}$ is regular, if and only if for any explosive eigenvalue $\lambda$, in the Jordan decomposition of $\trans{}$ there is only one block corresponding to $\lambda$. In other words, no matter how large the algebraic multiplicity of $\lambda$ is, its geometric multiplicity is one. Another equivalent formulation is the following one. $\trans{}$ is regular if and only if
\begin{eqnarray*}
	\rank{\trans{}-\lambda I_p} \geq p-1,
\end{eqnarray*}
for all $\lambda \in \C$ such that $\left|\lambda \right|>1$. For example, let $P_1,P_2 \in \C^{2 \times 2}$ be arbitrary invertible matrices, and
\begin{eqnarray*}
	\trans{1}=P_1^{-1}\begin{bmatrix}
		\rho & 1 \\ 0 & \rho
	\end{bmatrix} P_1, \:\:\: \trans{2} = P_2^{-1} \begin{bmatrix}
	\rho & 0 \\ 0 & \rho
\end{bmatrix} P_2,
\end{eqnarray*}
where $\rho \in \C, \left|\rho\right|>1$. Then, $\trans{1}$ is regular, where $\trans{2}$ is not.
\begin{propo} \label{minpoly}
	Assuming $\eigmin{\trans{0}}>1$, regularity of $\trans{0}$ is equivalent to $\minpoly{\trans{0}}>0$.
\end{propo}

The next proposition shows that positiveness of $\innerproductmin{\trans{0}}{\delta}$ is implied by reachability. Proposition \ref{PDcore} also reveals a linear scaling of $\innerproductmin{\trans{0}}{\delta}$ with respect to $\delta$, when the noise is a continuous random variable.

\begin{propo} \label{PDcore}
	Assume $\eigmin{\trans{0}}>1$, and $\left[\trans{0},C\right]$ is reachable. We then have $\innerproductmin{\trans{0}}{\delta}> 0$. Moreover, if there is $i \geq p$, such that $w(i-p+1), \cdots, w(i)$ have bounded probability density functions (pdf) over certain subspaces of $\R^p$, then, 
	\begin{eqnarray*}
		\innerproductmin{\trans{0}}{\delta} \geq \innerproductminconstant{\trans{0}} \delta,
	\end{eqnarray*}
	for some constant $\innerproductminconstant{\trans{0}} >0$. If the bounded pdfs mentioned above correspond to the normal distribution, then 
	\begin{eqnarray*}
		\innerproductminconstant{\trans{0}} \geq \left(\frac{\pi \eigmin{K(C)}}{2\eigmax{{\trans{0}}^{i}{\trans{0}'}^{i}}}\right)^{1/2} p^{-1} \left(\min\limits_{1 \leq i \leq p}\norm{P_i}{2}\right) .
	\end{eqnarray*} 
\end{propo}

Now, we are ready to state the key result for the time length required to achieve accurate estimation for an explosive transition matrix. 
\begin{corol} \label{explosiveestimation}
	Suppose that $\eigmin{\trans{0}}>1$, $\trans{0}$ is regular, and $\left[\trans{0},C\right]$ is reachable. There exists a constant $n_2 <\infty$, such that for all $\epsilon,\delta>0$,
	\edit{\begin{eqnarray} \label{explosiveestimationss}
		n \geq \frac{3 \left( \tailexp + 4 \right)}{\tailexp \log \eigmin{\trans{0}}} \log \left(\frac{- \log \delta}{\epsilon \innerproductmin{\trans{0}}{\delta}} \right) \vee n_2
	\end{eqnarray}}
	implies
	\begin{eqnarray*}
		\PP{\Mnorm{\esttrans{n}-\trans{0}}{2}>\epsilon} < 4\delta.
	\end{eqnarray*} 
\end{corol}

The time length specified in \eqref{explosiveestimationss} is similar to that of Lemma \ref{explosivemin} in terms of the accuracy $\epsilon$, while the dependence in $\delta$ is different. In fact, compared to \eqref{SSexplosivemin}, the decay rate of $\delta$ is of the common exponential order as $n$ grows (assuming the linear scaling of $\innerproductmin{\trans{0}}{\delta}$ with respect to $\delta$). 
\begin{remark}
	Another interesting property of explosive systems is that $n_1,n_2$ scale logarithmically with respect to the dimension $p$. 
\end{remark}
The dependency of $n_1,n_2$ on $\trans{0}$, as well as $\tailconst, \tailcoeff$, and $\tailexp$ specified in Assumption \ref{tailcondition} are outlined explicitly in the corresponding proofs. Moreover, the constants $n_1,n_2$ are in fact universal; for $\rho_1>0$, a single $n_1$ depending on $\rho_1$ implies \eqref{explosivemineq1} for all matrices $\trans{0}$ satisfying $\eigmin{\trans{0}} \geq 1+\rho_1$.

In addition, let $\lambda_1 \left(\trans{}\right), \cdots, \lambda_{k \left(\trans{}\right)} \left(\trans{}\right)$ be the distinct eigenvalues of $\trans{}$. Then, there is a single universal constant $n_2$ depending on $\rho_1,\rho_2$, such that \eqref{explosiveestimationss} implies the desired estimation result of Corollary \ref{explosiveestimation}, for all regular explosive transition matrices $\trans{0}$ satisfying $1 + \rho_1 \leq \eigmin{\trans{0}}$, and $0< \rho_2 \leq \min\limits_{1 \leq i < j \leq k \left( \trans{0} \right)} \left| \lambda_i\left( \trans{0} \right) - \lambda_j\left( \trans{0} \right) \right|$.
 
\subsection{General systems} \label{generalcase}
The previous results enable us to establish the key result of the paper. Theorem \ref{consistency} establishes the accuracy of identification, when the regular matrix $\trans{0}$ has no eigenvalue on the unit circle. \edit{As the following well known fact states, this assumption includes almost all matrices \cite{mumford1999red}.}
\begin{fact} \label{unitrootexclusion}
		The set of all $p \times p$ real matrices with at least one eigenvalue on the unit circle has Lebesgue measure zero. Moreover, almost all matrices are regular.
\end{fact}

\edit{However, note that transition matrices with unit eigenvalues occur in applications, including resonating mechanical systems \cite{fossen2011parametric}, the study of macroeconomic indicators \cite{engsted2006explosive,nelson1982trends}
and the timeline of bubbles during the crisis in the mid-late 2000s \cite{phillips2011dating}. Therefore, addressing the identification problem for unit root transition matrices, even though they constitute a measure zero set, is an interesting direction for future work.}

Excluding two pathological cases of square matrices with at least one eigenvalue on the unit circle, and irregular matrices, the estimation of the transition matrix for a general unstable system is with high probability arbitrarily accurate, as determined in the following theorem. A well known fact states that there is an invertible matrix $M \in \R^{p \times p}$, such that $\tilde{A}=M\trans{0}M^{-1} \in \R^{p \times p}$ is a block diagonal matrix,
\begin{eqnarray*}
	\tilde{A} = \begin{bmatrix}
		\trans{1} & 0 \\
		0 & \trans{2}
	\end{bmatrix},
\end{eqnarray*}
where for $i=1,2$, we have $\trans{i} \in \R^{p_i \times p_i}$, $p_1+p_2=p$, and
\begin{eqnarray*}
	\eigmax{\trans{1}}<1 < \eigmin{\trans{2}}.
\end{eqnarray*}
Technically, $p_1$ ($p_2$) is sum of the algebraic multiplicities of the stable (explosive) eigenvalues of the true unknown matrix $\trans{0}$. Conceptually, it determines the dimension of a certain subspace of $\R^p$, on which the linear transformation $\trans{0}$ is stable (explosive). Note that since $M$ is not known in advance, the above split of the true transition matrix to a stable one and an explosive one cannot be used in the identification procedure. 
\begin{thrm} \label{consistency}
	Suppose that $\trans{0}$ is regular, has no unit eigenvalue, $\left[\trans{0},C\right]$ is reachable, and $\trans{2}$ is as above. Then, there exist constants $\ssconstant_3, n_3 <\infty$, such that for all $\epsilon, \delta>0$, 
 	\edit{\begin{eqnarray} \label{generalcasecondition}
		\frac{n}{\left(\log n\right)^{4/\tailexp}} \geq \frac{\ssconstant_3}{\epsilon^2} \left( \left( -\log \delta \right)^{1+4/\tailexp} - \log \innerproductmin{\trans{2}}{\delta} \right ) \vee n_3, \:\:\:\:\:\:
	\end{eqnarray}}
	implies that
	\begin{eqnarray*}
		\PP{\Mnorm{\esttrans{n}-\trans{0}}{2}>\epsilon} < 6\delta.
	\end{eqnarray*}
\end{thrm}

\edit{Regarding the time length above in \eqref{generalcasecondition}, the exact specification of the constants $\ssconstant_3, n_3$ requires some additional definitions, provided in the proof of Theorem \ref{consistency}. Broadly speaking, the behavior of $\ssconstant_3$ (respectively $n_3$) is similar to that of $\ssconstant_2$ (resp. $n_2$) used in Corollary \ref{stableestimation} (resp. \ref{explosiveestimation}). Note that in order to compute $\ssconstant_3$ (resp. $n_3$), one has to use the stable (resp. explosive) matrix $\trans{1}$ (resp. $\trans{2}$). Further, since $\trans{0}$ is regular, regularity of $\trans{2}$ is automatically guaranteed. Therefore, Corollaries \ref{stableestimation} and \ref{explosiveestimation} can be used. Note that the reachability condition is inherited from the matrix $\trans{0}$, as formally presented in Proposition \ref{newmatricesreachability}. Thus, Proposition \ref{PDcore} implies that $-\log \innerproductmin{\trans{2}}{\delta} < \infty$, and it is up to a constant less than $-\log \delta$, if the noise vectors have bounded probability density functions. Therefore, using Proposition \ref{PDcore}, for continuously distributed noise vectors with bounded pdfs one can substitute \eqref{generalcasecondition} with
	\begin{eqnarray*}
		\frac{n}{\left(\log n\right)^{4/\tailexp}} \geq \frac{2\ssconstant_3}{\epsilon^2}  \left( -\log \delta \right)^{1+4/\tailexp} \vee \left(n_3 - 2 \log \innerproductminconstant{\trans{2}}\right). 
	\end{eqnarray*}
}

\section{Concluding Remarks}
We studied the problem of providing finite time bounds for the least-squares estimates of general linear dynamical systems, where the transition matrix does not necessarily need to be stable. The relationships between different parameters involved, including time length, accuracy of the identification, failure probability, the transition and noise matrices, and dimension are investigated. We prove that apart from a pathological case of zero Lebesgue measure, the identification is with high probability accurate, if the length of the time period scales similar to standard results in estimation theory, i.e. quadratic scaling with the inverse identification error and logarithmic scaling with the failure probability. 

These finite time results for such a widely used model can be helpful to obtain analogous results for more complicated models exhibiting temporal dependence, such as nonlinear systems. Further, the techniques used in this work to analyze the accuracy when the systems under study are not necessarily stable, provide insight for settings where additional knowledge on the structure of the dynamics is available. In particular, potential extensions to a high-dimensional setting (assuming that the transition matrix is sparse), or other structured classes such as low-rank matrices, \edit{as well as addressing practically interesting cases of null measure, are topics of interest and for future investigation.}





\bibliographystyle{IEEEtran}        
\bibliography{References}           

\appendix
\newpage

\section{Proofs of Main Results}
\subsection{\bf Proof of Lemma \ref{stablemin}}
In this proof, we use the following propositions.

\begin{propo} \label{noisebound}
	For $n=1,2,\cdots$, and $0<\delta<1$, define the following event.
	\begin{eqnarray*}
		\event{W} &=& \left\{ \max\limits_{1 \leq t \leq n} \norm{w(t)}{\infty} \leq \noisemax{n}{\delta} \right\}.	
	\end{eqnarray*}
	where $\noisemax{n}{\delta}=\left(\tailcoeff \log \frac{ \tailconst np}{\delta}\right)^{1/\tailexp}$. We have $\PP{\event{W}} \geq 1-\delta$.
\end{propo}

\begin{propo} \label{statenorm}
	If $\trans{0}$ is stable, on the event $\event{W}$ we have
	\begin{eqnarray*}
	\norm{x(t)}{2} \leq \MJordanconst{}{\trans{0}} \left(\norm{x(0)}{\infty} + \noisemax{n}{\delta} \right),
	\end{eqnarray*}
	for all $t=1,2,\cdots,n$.
\end{propo}
\begin{propo} \label{empiricalcov}
	Define $C_n=\frac{1}{n}\sum\limits_{i=1}^{n}w(i)w(i)'$, and assume \begin{eqnarray} \label{empiricalcoveq1}
	\frac{n}{\noisemax{n}{\delta}^2} \geq \frac{6 \eigmax{C}+2\epsilon}{3 \epsilon^2} p \log \left(\frac{2p}{\delta}\right).
	\end{eqnarray} 
	On the event $\event{W}$ we have
	\begin{eqnarray*}
	\PP{\eigmax{C_n-C}>\epsilon} \leq \delta.
	\end{eqnarray*} 
\end{propo}

\begin{propo} \label{crossproduct}
	For stable $\trans{0}$ define 
	\begin{eqnarray*}
		U_n &=& \frac{1}{n}\sum\limits_{i=0}^{n-1} \left[\trans{0}x(i)w(i+1)'+w(i+1)x(i)'\trans{0}'\right] , \\
		\predbound{n}{\delta} &=& \Mnorm{\trans{0}}{2} \MJordanconst{}{\trans{0}} \noisemax{n}{\delta} \left(\norm{x(0)}{\infty}  + \noisemax{n}{\delta} \right).
	\end{eqnarray*}
	Assuming 
	\begin{eqnarray*}
	\frac{n}{\predbound{n}{\delta}^2} \geq \frac{32 p}{\epsilon^2} \log \left(\frac{2p}{\delta}\right),
	\end{eqnarray*} 
	on the event $\event{W}$ we have
	\begin{eqnarray*}
	\PP{\eigmax{U_n}>\epsilon} \leq \delta.
	\end{eqnarray*} 
\end{propo}

	Next, letting $\predbound{n}{\delta}$ be the same as Proposition \ref{crossproduct}, suppose that $\samplesize{\ref{stablemin}}{\epsilon}{\delta}$ is large enough, such that \eqref{stablemineq1}, \eqref{stablemineq2}, and \eqref{stablemineq3} (next page) hold for all $n \geq \samplesize{\ref{stablemin}}{\epsilon}{\delta}$. 
	\begin{table*}
	\begin{eqnarray}
	\frac{n}{\noisemax{n}{\delta}^2} &\geq& \frac{18 \eigmax{C}+2\epsilon}{ \epsilon^2} p \MJordanconst{}{\trans{0}'}^4 \log \left(\frac{4p}{\delta}\right) ,\label{stablemineq1}\\
	\frac{n}{\predbound{n}{\delta}^2} &\geq& \frac{288 p}{\epsilon^2} \MJordanconst{}{\trans{0}'}^4 \log \left(\frac{4p}{\delta}\right) ,\label{stablemineq2}\\
	\frac{n}{\left(\norm{x(0)}{\infty} + \noisemax{n}{\delta} \right)^2} &\geq& \frac{6}{\epsilon} \left(\Mnorm{\trans{0}}{2}^2+1\right) \MJordanconst{}{\trans{0}'}^2 \MJordanconst{}{\trans{0}}^2 .\label{stablemineq3}
	\end{eqnarray}
	\end{table*}
We prove that on the event $\event{W}$, for all $n \geq \samplesize{\ref{stablemin}}{\epsilon}{\delta}$ we have 
	\begin{eqnarray*}
	\PP{\eigmax{\frac{1}{n}V_{n+1}- \lyap{C}}>\epsilon} < \delta.
	\end{eqnarray*}
	First, according to \eqref{VARdynamics} we have
	\begin{eqnarray*}
		V_{n+1} &=& x(0)x(0)'+\trans{0}\sum\limits_{i=0}^{n-1}x(i)x(i)'\trans{0}'+ \sum\limits_{i=1}^{n} w(i) w(i)'  \\
		&+& \sum\limits_{i=0}^{n-1} \left[\trans{0}x(i)w(i+1)'+w(i+1)x(i)'\trans{0}'\right] \\
		&=& \trans{0}V_{n+1}\trans{0}'+nU_n+nC_n \\
		&+& \trans{0}\left(x(0)x(0)' - x(n)x(n)'\right)\trans{0}'+x(0)x(0)',
	\end{eqnarray*}
	where $C_n$, and $U_n$ are defined in Proposition \ref{empiricalcov}, and Proposition \ref{crossproduct}, respectively. Letting $E_n=U_n+C_n+ \frac{1}{n} \trans{0}\left(x(0)x(0)' - x(n)x(n)'\right)\trans{0}'+\frac{1}{n}x(0)x(0)'$, since $\eigmax{\trans{0}}<1$, the Lyapunov equation $V_{n+1}=\trans{0}V_{n+1}\trans{0}'+nE_n$ has the solution
	\begin{eqnarray*}
	\frac{1}{n}V_{n+1} = \sum\limits_{i=0}^{\infty}\trans{0}^i E_n {\trans{0}'}^i = \lyap{E_n}.
	\end{eqnarray*}
	Henceforth in the proof, we assume the event $\event{W}$ holds. According to Proposition \ref{empiricalcov}, \eqref{stablemineq1} implies that
	\begin{eqnarray} \label{stablemineproof1}
	\PP{\eigmax{C_n-C}>\frac{\epsilon}{3 \MJordanconst{}{\trans{0}'}^2}} \leq \frac{\delta}{2}.
	\end{eqnarray}
	In addition, by Proposition \ref{crossproduct}, \eqref{stablemineq2} implies that
	\begin{eqnarray} \label{stablemineproof2}
	\PP{\eigmax{U_n}>\frac{\epsilon}{3\MJordanconst{}{\trans{0}'}^2}} \leq \frac{\delta}{2}.
	\end{eqnarray} 
	Finally, using Proposition \ref{statenorm}, by \eqref{stablemineq3} we get
	\begin{eqnarray} \label{stablemineproof3}
	\frac{1}{n}\left(\Mnorm{\trans{0}}{2}^2 +1\right) \left( \norm{x(0)}{2}^2 + \norm{x(n)}{2}^2\right) \leq \frac{\epsilon}{3\MJordanconst{}{\trans{0}'}^2}.
	\end{eqnarray}
	Now, similar to the proof of Proposition \ref{statenorm}, we have $\sum\limits_{t=0}^{\infty} \Opnorm{{\trans{0}'}^t}{\infty}{2} \leq \MJordanconst{}{\trans{0}'}$, which because of $ \Mnorm{{\trans{0}'}^t}{2} \leq  \Opnorm{{\trans{0}'}^t}{\infty}{2}$ leads to
	\begin{eqnarray} \label{stablemineproof4}
	\sum\limits_{t=0}^{\infty} \Mnorm{{\trans{0}'}^t}{2}^2 \leq \left(\sum\limits_{t=0}^{\infty} \Mnorm{{\trans{0}'}^t}{2}\right)^2 \leq \MJordanconst{}{\trans{0}'}^2.
	\end{eqnarray}
	Putting \eqref{stablemineproof1}, \eqref{stablemineproof2}, \eqref{stablemineproof3}, and \eqref{stablemineproof4} together, on the event $\event{W}$ we have  
	\begin{eqnarray*} 
		\eigmax{\lyap{E_n-C}} &\leq& \sum\limits_{t=0}^{\infty} \eigmax{\trans{0}^t \left(E_n - C\right){\trans{0}'}^t} \\
		&\leq& \MJordanconst{}{\trans{0}'}^2 \eigmax{E_n-C} \leq \epsilon,
	\end{eqnarray*}
	with probability at least $1-\delta$. Then, since the definition of $\ssconstant_1$ implies \eqref{stableestimationeq1} - \eqref{stableestimationeq3} for $\delta/2$, the desired result holds.

\endproof

\subsection{\bf Proof of Corollary \ref{stableestimation}}
	We prove that if the followings hold, then on the event $\event{W}$ we have $\Mnorm{\esttrans{n}-\trans{0}}{2} \leq \epsilon$, with probability at least $1-\delta$. Letting $\samplesize{\ref{stablemin}}{\cdot}{\cdot}$ be as defined in the proof of Lemma \ref{stablemin}, suppose that
	\begin{eqnarray}
	n &\geq& \samplesize{\ref{stablemin}}{\frac{\eigmin{K(C)}}{2}}{\frac{\delta}{2}}+1, \label{stableestimationeq1} \\
	\frac{n-2}{\predbound{n}{\delta}^2} &\geq& \frac{32p}{\eigmin{K(C)}^2\epsilon^2} \log \left(\frac{4p}{\delta}\right) ,\label{stableestimationeq2}
	\end{eqnarray}
	where 
	\begin{eqnarray*}
	\predbound{n}{\delta} = \MJordanconst{}{\trans{0}} \noisemax{n}{\delta} \left(\norm{x(0)}{\infty}  + \noisemax{n}{\delta} \right).
	\end{eqnarray*}
	First, by Lemma \ref{stablemin}, \eqref{stableestimationeq1} implies that on the event $\event{W}$,
	\begin{eqnarray} \label{stableestimationeq3}
	\frac{\eigmin{V_n}}{n-1}
	&\geq& \frac{\eigmin{K(C)}}{2},
	\end{eqnarray} 
	with probability at least $1-\delta/2$. Since $\left[\trans{0},C\right]$ is reachable, $\eigmin{K(C)}>0$. Thus, 
	\begin{eqnarray*}
	\esttrans{n}= \sum\limits_{t=0}^{n-1}x(t+1)x(t)' V_n^{-1} = \trans{0}+U_n V_n^{-1}, 
	\end{eqnarray*}
	where $U_n=\sum\limits_{t=0}^{n-1}w(t+1)x(t)'$, which leads to 
	\begin{eqnarray} \label{stableestimationeproofeq1}
	\Mnorm{\esttrans{n}-\trans{0}}{2} \leq \frac{\Mnorm{U_n}{2}}{\eigmin{V_n}}.
	\end{eqnarray}
	To proceed, for an arbitrary matrix $H \in \R^{k \times \ell}$, defining the linear transformation 
	\begin{eqnarray*}
	\symmetrizer{H} = \begin{bmatrix} 0_{k \times k} & H \\ H' & 0_{\ell \times \ell} \end{bmatrix} \in \R^{(k+\ell) \times (k+\ell)},
	\end{eqnarray*}
	it holds that $\Mnorm{H}{2} = \eigmax{\symmetrizer{H}}$ (see \cite{tropp2012user}). Note that $\symmetrizer{H}$ is always symmetric. Next, letting $X_t=w(t+1)x(t)'$, apply Proposition \ref{MAzuma} to $\symmetrizer{X_t} \in \R^{2p \times 2p}$. Since
	\begin{eqnarray*}
	\symmetrizer{X_t}^2 = \begin{bmatrix} \norm{x(t)}{2}^2 w(t+1)w(t+1)' & 0_{p\times p} \\ 0_{p \times p} & \norm{w(t+1)}{2}^2 x(t)x(t)' \end{bmatrix},
	\end{eqnarray*}
	by Proposition \ref{noisebound}, and Proposition \ref{statenorm}, all matrices $\symmetrizer{M_t}^2-\symmetrizer{X_t}^2$ are positive semidefinite on the event $\event{W}$, where
	\begin{eqnarray*}
	M_t= p^{1/2} \MJordanconst{}{\trans{0}} \noisemax{n}{\delta} \left(\norm{x(0)}{\infty}  + \noisemax{n}{\delta} \right) I_p.
	\end{eqnarray*}
	By
	\begin{eqnarray*}
	\sigma^2=\eigmax{\sum\limits_{t=0}^{n-1} \symmetrizer{M_t}^2}= np \predbound{n}{\delta}^2,
	\end{eqnarray*}
	letting $y=\frac{\eigmin{K(C)}}{2}\left(n-1\right) \epsilon$, according to Proposition \ref{MAzuma}, \eqref{stableestimationeq2} implies 
	\begin{eqnarray*}
	\PP{\Mnorm{U_n}{2}>y} = \PP{\eigmax{\symmetrizer{U_n}}>y} \leq \frac{\delta}{2},
	\end{eqnarray*}
	which in addition to \eqref{stableestimationeq3} gives the desired result, once plugged in \eqref{stableestimationeproofeq1}. \\Finally, the definition of $\ssconstant_2$ implies all above statements for $\delta/2$, which completes the proof.
	
\endproof

\subsection{\bf Proof of Lemma \ref{explosivemin}}

\begin{propo} \label{zinfinitynormbound}
	Let $z(n)=x(0)+ \sum\limits_{t=1}^{n} \trans{0}^{-t} w(t)$, where $\trans{0}$ is an explosive matrix with Jordan decomposition $\trans{0}=P^{-1}\Lambda P$. Define the event
	\begin{eqnarray*}
	\event{V} = \left\{\sup\limits_{1 \leq n \leq \infty} \norm{z(n)}{2} \leq \zinfinitybound{\trans{0}}{\delta} \right\} ,
	\end{eqnarray*} 
	where $\zinfinitybound{\trans{0}}{\delta}$ is defined in \eqref{zinfinitybounddef} (next page). Then, we have $\PP{\event{V}} \geq 1-\delta$. 
\end{propo}

	Letting $\trans{0}=P^{-1} \Lambda P$ be the Jordan decomposition, and $z(0)=x(0)$, for $n=1,2, \cdots$, define
	\begin{eqnarray*}
	z(n) &=& x(0)+ \sum\limits_{t=1}^{n} \trans{0}^{-t} w(t),\\ 
	U_n &=& \trans{0}^{-n}V_{n+1} {\trans{0}'}^{-n}, \\
	F_n &=& \sum\limits_{t=0}^{n} \trans{0}^{-t} z(n)z(n)' {\trans{0}'}^{-t}.
	\end{eqnarray*}
	First, using $x(t)=\trans{0}^t z(t)$, since
	\begin{eqnarray*}
		U_n = \sum\limits_{t=0}^{n} \trans{0}^{-n} x(t)x(t)' {\trans{0}'}^{-n} = \sum\limits_{t=0}^{n} \trans{0}^{-n+t} z(t)z(t)' {\trans{0}'}^{-n+t} ,
	\end{eqnarray*}
	by Proposition \ref{zinfinitynormbound}, on the event $\event{V}$ we have
	\begin{eqnarray*} 
	&&\eigmax{U_n} \leq \sum\limits_{t=0}^{\infty} \norm{\trans{0}^{-t} z(n-t)}{2}^2 \\ &\leq& \sum\limits_{t=0}^{\infty} \Mnorm{\trans{0}^{-t}}{2}^2 \norm{z(n-t)}{2}^2 \leq \MJordanconst{}{\trans{0}^{-1}}^2 \zinfinitybound{\trans{0}}{\delta}^2 ,
	\end{eqnarray*}
	which is the desired result, because the right hand side above is at most $ \zinfinityconstant{\trans{0}} \left(- \log \delta \right)^{2/\tailexp}$, for the constant $\zinfinityconstant{\trans{0}}$ defined by \eqref{zinfinityconstantdef} (next page). In the sequel, we prove the desired result about the smallest eigenvalue. Letting $\rho_1,\rho_2$ be as defined in \eqref{rho1define}, \eqref{rho2define} (next page), 
	\begin{table*}
		\begin{eqnarray}
		\zinfinitybound{\trans{0}}{\delta} &=& \norm{x(0)}{2} + \Opnorm{P^{-1}}{\infty}{2} \Mnorm{P}{\infty} \sum\limits_{t=1}^{\infty} \MJordanconst{t}{\Lambda^{-1}} \left(\tailcoeff \log \frac{2 \tailconst p t^2}{\delta}\right)^{1/\tailexp} < \infty, \label{zinfinitybounddef}\\
		\zinfinityconstant{\trans{0}} &=& \MJordanconst{}{\trans{0}^{-1}}^2\left[\norm{x(0)}{2} + \Opnorm{P^{-1}}{\infty}{2} \Mnorm{P}{\infty} \sum\limits_{t=1}^{\infty} \MJordanconst{t}{\Lambda^{-1}} \tailcoeff^{1/\tailexp} \log \left(2 \tailconst p t^2\right)^{1/\tailexp}\right]^2, \label{zinfinityconstantdef} \\
		\rho_1 &=& 2 \left(\Opnorm{P^{-1}}{\infty}{2} \Mnorm{P}{\infty}  \MJordanconst{}{\trans{0}'^{-1}}^2 + \MJordanconst{}{\trans{0}^{-1}} \Opnorm{P'}{\infty}{2}^2 \Mnorm{{P'}^{-1}}{\infty}^2\right) e^{2\eigmin{\trans{0}}}, \label{rho1define}\\
		\rho_2 &=& 2 \MJordanconst{}{\trans{0}'^{-1}}^2 \left(2 + \MJordanconst{}{\trans{0}^{-1}}\right) \Opnorm{P^{-1}}{\infty}{2} \Mnorm{P}{\infty} e^{\eigmin{\trans{0}}}. \label{rho2define}
		\end{eqnarray}
	\end{table*}
	assume the followings hold for all $n \geq \samplesize{\ref{explosivemin}}{\epsilon}{\delta}$:
	\begin{eqnarray}
	\noisemax{n}{\delta} n^{2\mult{\trans{0}}} \eigmin{\trans{0}}^{-2n/3} &\leq& \frac{\epsilon}{\rho_1 \zinfinitybound{\trans{0}}{\delta}} , \label{explosivemincondition1}\\
	n^{\mult{\trans{0}}-1} \eigmin{\trans{0}}^{-n} &\leq& \frac{\epsilon}{\rho_2 \zinfinitybound{\trans{0}}{\delta}^2} \label{explosivemincondition2}, 
	\end{eqnarray}
	where $\mult{\trans{0}}$ is defined after Definition \ref{Jordandeff}. \edit{Note that taking
	\begin{eqnarray*}
	n_1 = \frac{3\log \left( \rho_1 \rho_2 \zinfinityconstant{\trans{0}}^3 \left( \tailcoeff \log \tailconst p \right)^{1/\tailexp} \right)+ 12 \mult{\trans{0}}+ 6/\tailexp}{\eigmin{\trans{0}}} ,
	\end{eqnarray*}
	\eqref{SSexplosivemin} implies \eqref{explosivemincondition1} and \eqref{explosivemincondition2}.}
	
	For all $n \geq \samplesize{\ref{explosivemin}}{\epsilon}{\delta}$, we show that with probability at least $1-4\delta$ it holds that
	\begin{eqnarray*}
	\eigmin{\trans{0}^{-n} V_{n+1} {\trans{0}'}^{-n}} < \minpoly{\trans{0}}^2 \innerproductmin{\trans{0}}{\delta}^2 - \epsilon.
	\end{eqnarray*} 
	The proof is based on the following propositions.	
	\begin{propo} \label{explosiveminaux1}
		On the event $\event{W} \cap \event{V}$, we have
		\begin{eqnarray} \label{explosiveminproofeq2}
		\eigmax{U_n - F_n} \leq \frac{\epsilon}{2}.
		\end{eqnarray}
	\end{propo}

	\begin{propo} \label{explosiveminaux2}
		On $\event{V}$, with probability at least $1-\delta$,
		\begin{eqnarray} \label{explosiveminproofeq3}
		\eigmax{F_\infty - F_n} \leq \frac{\epsilon}{2}.
		\end{eqnarray}
	\end{propo}
	Next, we show that with probability at least $1-\delta$,
	\begin{eqnarray} \label{explosiveminproofeq5}
	\eigmin{F_\infty} \geq \left(\minpoly{\trans{0}}\innerproductmin{\trans{0}}{\delta}\right)^2 = \lambda_0.
	\end{eqnarray}
	For this purpose, we need the following propositions.
	\begin{propo} \label{explosiveminauxiliary}
		Letting $f(x)=\sum\limits_{i=0}^{p-1}a_{i+1}x^i$ be a real polynomial, we have
		\begin{eqnarray*}
		\PP{\norm{f\left(\trans{0}^{-1}\right) z(\infty)}{2} < \norm{a}{1} \minpoly{\trans{0}} \innerproductmin{\trans{0}}{\delta}} \leq \delta.
		\end{eqnarray*}
	\end{propo}
	\begin{propo} \label{zinfinityfullrankness}
		If $\minpoly{\trans{0}} \innerproductmin{\trans{0}}{\delta} \neq 0$, then,
		\begin{eqnarray*}
		\PP{\rank{\left[z(\infty), \trans{0}z(\infty), \cdots, \trans{0}^{-p+1}z(\infty)\right]}<p}=0.
		\end{eqnarray*}
	\end{propo}
	If $\lambda_0=0$, \eqref{explosiveminproofeq5} is trivial. Otherwise, assume $\eigmin{F_\infty}< \lambda_0$, and let $v \in \R^p$ be such that $\norm{v}{2}=1$, and $v'F_\infty v < \lambda_0$. Then,
	\begin{eqnarray*}
		&& \lambda_0 > \sum\limits_{t=0}^{p-1} v' \trans{0}^{-t} z(\infty)z(\infty)' {\trans{0}'}^{-t} v \\
		&\geq& \norm{v' \left[ z(\infty), \cdots , \trans{0}^{-p+1} z(\infty)\right]}{\infty}^2 = \max\limits_{0 \leq i \leq p-1} \left|v' \trans{0}^{-i} z(\infty)\right|^2 .
	\end{eqnarray*}
	By Proposition \ref{zinfinityfullrankness}, almost surely, there is $a \in \R^p$, such that $v=\sum\limits_{i=0}^{p-1} a_{i+1} \trans{0}^{-i}z(\infty)$. So,
	\begin{eqnarray*}
	\norm{v}{2} &=& \left|v' \left(\sum\limits_{i=0}^{p-1} a_{i+1} \trans{0}^i\right) z(\infty) \right| \leq \sum\limits_{i=0}^{p-1} \left|a_{i+1}\right| \left|v' \trans{0}^i z(\infty) \right| \\
	&<& \sum\limits_{i=0}^{p-1} \left|a_{i+1}\right| {\lambda_0}^{1/2} = {\lambda_0}^{1/2} \norm{a}{1},
	\end{eqnarray*}
	which, by Proposition \ref{explosiveminauxiliary}, holds with probability at most $\delta$. Putting \eqref{explosiveminproofeq2}, \eqref{explosiveminproofeq3}, and \eqref{explosiveminproofeq5} together, on the event $\event{W} \cap \event{V}$, we get the following, which holds with probability at least $1-2\delta$:
	\begin{eqnarray*}
		\eigmin{U_n} \geq \lambda_0 - \epsilon,
	\end{eqnarray*}
	which is the desired result.
\endproof

\subsection{\bf Proof of Corollary \ref{explosiveestimation}}
	Indeed, we prove that if the followings hold, then, we have $\Mnorm{\esttrans{n}-\trans{0}}{2} \leq \epsilon$, with probability at least $1-4\delta$. Letting $\samplesize{\ref{explosivemin}}{\cdot}{\cdot}$ be as defined in the proof of Lemma \ref{explosivemin}, suppose that
	\begin{eqnarray}
	n &\geq& \samplesize{\ref{explosivemin}}{\lambda_0}{\delta} +1 , \label{explosiveestimationeq1} \\
	\lambda_0 \epsilon &\geq& \rho \noisemax{n}{\delta} n^{\mult{\trans{0}}-1} \eigmin{\trans{0}}^{-n+1} , \label{explosiveestimationeq2} 
	\end{eqnarray}
	where 
	\begin{eqnarray*}
	\lambda_0 &=& \frac{1}{2} \minpoly{\trans{0}}^2 \innerproductmin{\trans{0}}{\delta}^2, \\ 
	\rho &=& p^{1/2} \zinfinitybound{\trans{0}}{\delta} \MJordanconst{}{\trans{0}^{-1}} \Opnorm{P^{-1}}{\infty}{2} \Mnorm{P}{\infty} e^{\eigmin{\trans{0}}}.
	\end{eqnarray*}
	\edit{Note that taking
	\begin{eqnarray*}
	n_2 &=& n_1 + \frac{3}{\eigmin{\trans{0}}} \log \left(\frac{2p \zinfinityconstant{\trans{0}} \MJordanconst{}{\trans{0}^{-1}}}{ \minpoly{\trans{0}}} \right) \\
	&+& \frac{3}{\eigmin{\trans{0}}} \log \left( \Opnorm{P^{-1}}{\infty}{2} \Mnorm{P}{\infty} e^{\eigmin{\trans{0}}} \right),
	\end{eqnarray*}
	\eqref{explosiveestimationss} implies \eqref{explosiveestimationeq1}, \eqref{explosiveestimationeq2}.}
	
	First, by Lemma \ref{explosivemin}, \eqref{explosiveestimationeq1} implies that on the event $\event{W} \cap \event{V}$,
	\begin{eqnarray} \label{explosiveestimationeq3}
	\eigmin{\trans{0}^{-n+1} V_{n} {\trans{0}'}^{-n+1}} \geq \lambda_0,
	\end{eqnarray} 
	with probability at least $1-2\delta$. According to Proposition \ref{minpoly} and Proposition \ref{PDcore}, regularity, in addition to reachability, imply $\lambda_0 > 0$. Thus, 
	\begin{eqnarray*}
	\esttrans{n}= \sum\limits_{t=0}^{n-1}x(t+1)x(t)' V_n^{-1} = \trans{0}+U_n {\trans{0}'}^{n-1} V_n^{-1} , 
	\end{eqnarray*}
	where $U_n=\sum\limits_{t=0}^{n-1}w(t+1)x(t)'{\trans{0}'}^{-n+1} $, which leads to 
	\begin{eqnarray} \label{explosiveestimationeq4}
	\Mnorm{\esttrans{n}-\trans{0}}{2} \leq \frac{\Mnorm{U_n}{2} \Mnorm{{\trans{0}}^{-n+1} }{2}}{\eigmin{{\trans{0}}^{-n+1} V_n {\trans{0}}^{-n+1} }}.
	\end{eqnarray}
	Since $x(t)=\trans{0}^t z(t)$, Proposition \ref{noisebound} and Proposition \ref{zinfinitynormbound} imply that on the event $\event{W} \cap \event{V}$,
	\begin{eqnarray} \label{explosiveestimationeq5}
	\Mnorm{U_n}{2} \leq p^{1/2}\noisemax{n}{\delta} \zinfinitybound{\trans{0}}{\delta} \MJordanconst{}{\trans{0}^{-1}}
	\end{eqnarray}
	Plugging \eqref{explosiveestimationeq3} and \eqref{explosiveestimationeq5} in \eqref{explosiveestimationeq4}, and using \eqref{Jordaneq1}, we get
	\begin{eqnarray*}
	\Mnorm{\esttrans{n}-\trans{0}}{2} \leq \frac{\rho}{\lambda_0} \noisemax{n}{\delta} n^{\mult{\trans{0}}-1} \eigmin{\trans{0}}^{n-1},
	\end{eqnarray*}
	which by \eqref{explosiveestimationeq2} is at most $\epsilon$, holding with probability at least $1-2\delta$ on $\event{W} \cap \event{V}$.
\endproof

\subsection{\bf Proof of Theorem \ref{consistency}}
	We split the original system into two parts, each with transition matrix $\trans{i}$. First, let
	\begin{eqnarray*}
	\tilde{C}=MCM'= \begin{bmatrix} C_{11} & C_{12} \\ C_{21} & C_{22} \end{bmatrix},
	\end{eqnarray*}
	where $C_{ij} \in \R^{p_i \times p_j}$ for $i=1,2$. Then, for $t=0,1,\cdots$, defining 
	\begin{eqnarray*}
	\tilde{x}(t) &=& Mx(t), \\ 
	\tilde{w}(t+1) &=& Mw(t+1),
	\end{eqnarray*}
	we have
	\begin{eqnarray*}
	\tilde{x}(t+1) &=& M \left( \trans{0} x(t)+w(t+1)\right) \\
	&=& \tilde{A} M x(t)+ M w(t+1) \\
	&=& \tilde{A}\tilde{x}(t) + \tilde{w}(t+1). 
	\end{eqnarray*}
	Note that letting 
	\begin{eqnarray*}
	\noisemax{n+1}{\delta}=\left( \Mnorm{M}{\infty} \vee 1 \right) \left(\tailcoeff \log \frac{ \tailconst p \left(n+1\right)}{\delta}\right)^{1/\tailexp},
	\end{eqnarray*}
	similar to Proposition \ref{noisebound}, we have $\PP{\event{W}} \geq 1-\delta$, where 
	\begin{eqnarray*}
	\event{W} = \left\{ \max\limits_{1 \leq t \leq n+1}  \left( \norm{w(t)}{\infty} \vee \norm{\tilde{w}(t)}{\infty}\right) \leq \noisemax{n+1}{\delta} \right\}.
	\end{eqnarray*}
	Let
	\begin{eqnarray*}
	\tilde{x}(t) &=& \left[x^{(1)}(t)', x^{(2)}(t)'\right]', \\
	\tilde{w}(t+1) &=& \left[w^{(1)}(t+1)', w^{(2)}(t+1)'\right]',
	\end{eqnarray*}
	where $x^{(i)}(t),w^{(i)}(t+1) \in \R^{p_i}$, for $i=1,2$. Since $\tilde{A}$ is block diagonal, the processes $x^{(1)}(t), x^{(2)}(t)$ are separated:
	\begin{eqnarray*}
		x^{(i)}(t+1) &=& \trans{i} x^{(i)}(t) + w^{(i)}(t+1), \\
		C_{ii} &=& \E{w^{(i)}(t+1)w^{(i)}(t+1)'}.
	\end{eqnarray*}
	Both new processes inherit reachability from the original one.
	\begin{propo} \label{newmatricesreachability}
		If $\left[\trans{0},C\right]$ is reachable, then for $i=1,2$, $\left[\trans{i},C_{ii}\right]$ is reachable as well.
	\end{propo}
	Now, we define the following parameters, which will be used in the proof. Letting $\trans{2}=P^{-1} \Lambda_2 P$ be the Jordan decomposition of the explosive matrix $\trans{2}$, and $K_1=\sum\limits_{t=0}^{\infty} \trans{1}^t C_{11} {\trans{1}'}^t$, define
	\begin{eqnarray*}
		\rho_0 &=& \frac{1}{2} - \frac{1}{2} \left(1- \frac{ \eigmin{K_1}}{9\eigmax{K_1}}\right)^{1/2}, \\
		\rho_1 &=& \frac{2 p \eigmin{\trans{2}} \zinfinitybound{\trans{2}}{\delta} \Opnorm{P'}{\infty}{2} \Mnorm{{P'}^{-1}}{\infty} e^{\eigmin{\trans{2}}} }{\minpoly{\trans{2}}\innerproductmin{\trans{2}}{\delta}} , \\
		\rho_2 &=& \frac{8 \MJordanconst{}{{\trans{2}'}^{-1}}^2 \zinfinitybound{\trans{2}}{\delta} \Opnorm{P^{-1}}{\infty}{2} \Mnorm{P}{\infty} e^{\eigmin{\trans{2}}}}{ \minpoly{\trans{2}}^{2} \innerproductmin{\trans{2}}{\delta}^{2}} , \\
		\rho_3 &=& \frac{4 \left(4 \eigmin{K_1}^{-1} +3\right)^{1/2} \Mnorm{M}{2}}{\eigmin{K_1}^{1/2} \rho_0},\\
		\rho_4 &=& \frac{ 2 p^{1/2} \zinfinitybound{\trans{2}}{\delta} \Opnorm{P^{-1}}{\infty}{2} \Mnorm{P}{\infty} e^{\eigmin{\trans{2}}}}{ \minpoly{\trans{2}} \innerproductmin{\trans{2}}{\delta}}  , \\
		\rho_5 &=& \frac{2 \Opnorm{P^{-1}}{\infty}{2} \Mnorm{P}{\infty} e^{\eigmin{\trans{2}}}}{\minpoly{\trans{2}} \innerproductmin{\trans{2}}{\delta}} , \\
		\rho_6 &=& \frac{ \Opnorm{P'}{\infty}{2} \Mnorm{P'^{-1}}{\infty} e^{\eigmin{\trans{2}}}\eigmin{K_1}^{1/2}}{\minpoly{\trans{2}} \innerproductmin{\trans{2}}{\delta} }.
	\end{eqnarray*}
	Note that the constants $\rho_0, \rho_3$ do not depend on $\delta$, and all other parameters depend on $\delta$, only through $\zinfinitybound{\trans{0}}{\delta}$ and $\innerproductmin{\trans{0}}{\delta}$. Using $\samplesize{\ref{stablemin}}{\cdot}{\cdot}$, and $\samplesize{\ref{explosivemin}}{\cdot}{\cdot}$ defined in Lemma \ref{stablemin} and Lemma \ref{explosivemin}, respectively, suppose that the conditions \eqref{consistencycondition1} - \eqref{consistencycondition7} (next page) hold.
	
	\begin{table*}
		\begin{eqnarray}
		n &\geq& \samplesize{\ref{explosivemin}}{\frac{\minpoly{\trans{2}}^2 \innerproductmin{\trans{2}}{\delta}^2}{2}}{\delta} \label{consistencycondition1} , \\
		n &\geq& 3 \samplesize{\ref{stablemin}}{\frac{\eigmin{K_1}}{2}}{\delta} \label{consistencycondition2} ,\\
		\frac{\rho_0}{\rho_1} &\geq& n^{\mult{\trans{2}}-1/2} \eigmin{\trans{2}}^{-2n/3} , \label{consistencycondition3} \\
		\frac{1}{\rho_2} &\geq& \noisemax{n+1}{\delta} n^{\mult{\trans{2}}} \eigmin{\trans{2}}^{-n/3} , \label{consistencycondition4} \\
		\frac{\epsilon}{3 \rho_3 \rho_4} 
		&\geq& \noisemax{n+1}{\delta} n^{\mult{\trans{2}}-1/2} \eigmin{\trans{2}}^{-2n/3} , \label{consistencycondition6} \\
		\frac{\epsilon}{3 \rho_3 \rho_5} &\geq& \noisemax{n+1}{\delta}^2 n^{\mult{\trans{2}}+1/2} \eigmin{\trans{2}}^{-n/3} , \label{consistencycondition8} \\
		\frac{1}{\rho_6} &\geq& n^{\mult{\trans{2}}-1/2} \eigmin{\trans{2}}^{-n} , \label{consistencycondition9} \\
		\frac{n^2 (n+1)^{-1}}{\left(\norm{x^{(1)}(0)}{\infty}+ \noisemax{n+1}{\delta}\right)^2 \noisemax{n+1}{\delta}^2} &\geq& \frac{8 p \rho_3^2 \MJordanconst{}{\trans{1}}^2}{\epsilon^2} \log \left(\frac{4 \left(p+p_1\right)}{\delta}\right) , \label{consistencycondition5} \\
		\frac{n}{\noisemax{n+1}{\delta}^2} &\geq& \frac{72 p^2 \rho_3^2}{\epsilon^2} \log \left(\frac{4 \left(p+p_2\right)}{\delta}\right). \label{consistencycondition7}
		\end{eqnarray} 
	\end{table*}
	We show that $\Mnorm{\esttrans{n+1}-\trans{0}}{2} \leq \epsilon$, with probability at least $1-6 \delta$.
	
	\edit{Among the conditions \eqref{consistencycondition1} - \eqref{consistencycondition7}, the main inequalities \\
		for $\epsilon$ are \eqref{consistencycondition5}, \eqref{consistencycondition7}, \\
		for $\innerproductmin{\trans{2}}{\delta}$ are \eqref{consistencycondition1}, \eqref{consistencycondition3}, \eqref{consistencycondition4}, \eqref{consistencycondition6}, \eqref{consistencycondition8}, \eqref{consistencycondition9},\\
		and for $\delta$ are \eqref{consistencycondition2}, \eqref{consistencycondition5}, \eqref{consistencycondition7}.\\
		Therefore, taking
	\begin{eqnarray}
	\ssconstant_3 &=& 72 p \left(\Mnorm{M}{2} \vee 1 \right)^4 \rho_3^2 \ssconstant_2 + \frac{18 \left( \tailexp+4 \right)}{\tailexp \log \eigmin{\trans{2}}}, \\
	n_3 &=& 12 \left(n_2 + \log \left( \rho_3 \Mnorm{M}{\infty} \vee 1 \right) \right),
	\end{eqnarray}
	\eqref{generalcasecondition} implies \eqref{consistencycondition1} - \eqref{consistencycondition7}. Above $\ssconstant_2$ is computed for the pair of matrices $\left[\trans{1}, C_{11}\right]$, and $n_2$ is computed for the pair $\left[\trans{2}, C_{22}\right]$.}
	
	First, 
	\begin{eqnarray*}
	&& MV_{n+1}M' = \sum\limits_{t=0}^{n} \tilde{x}(t)\tilde{x}(t)' \\
	&=& \sum\limits_{t=0}^{n} \begin{bmatrix} x^{(1)}(t) \\ x^{(2)}(t) \end{bmatrix} \left[x^{(1)}(t)' , x^{(2)}(t)'\right]= \begin{bmatrix} V_{n+1}^{(1)} & Y_{n+1}' \\ Y_{n+1} & V_{n+1}^{(2)} \end{bmatrix},
	\end{eqnarray*}
	where for $i=1,2$,
	\begin{eqnarray*}
		V_n^{(i)} &=& \sum\limits_{t=0}^{n-1} x^{(i)}(t)x^{(i)}(t)', \\
		Y_n &=& \sum\limits_{t=0}^{n-1} x^{(2)}(t)x^{(1)}(t)'.
	\end{eqnarray*}
	Let the event $\event{E} \subset \event{W} \cap \event{V}$ be the following:
	\begin{eqnarray*}
		\eigmin{\frac{1}{n}V_{n+1}^{(1)}} &\geq& \frac{1}{2} \eigmin{K_1}, \\
		\eigmin{\trans{2}^{-n} V^{(2)}_{n+1} {\trans{2}'}^{-n}} &\geq& \frac{1}{2} \minpoly{\trans{2}}^2 \innerproductmin{\trans{2}}{\delta}^2.
	\end{eqnarray*} 
	According to Lemma \ref{stablemin}, and Lemma \ref{explosivemin}, \eqref{consistencycondition1}, \eqref{consistencycondition2} imply $\PP{\event{E}} > 1-5\delta$. Henceforth in the proof, we assume the event $\event{E}$ holds. Define the invertible symmetric matrix 
	\begin{eqnarray*}
	U_n = \begin{bmatrix} {V_{n+1}^{(1)}} & 0_{p_1 \times p_2} \\ 0_{p_2 \times p_1} & {V_{n+1}^{(2)}} \end{bmatrix}^{-1/2} \in \R^{p \times p} ,
	\end{eqnarray*} 
	and let
	\begin{eqnarray*}
	&& E_n = U_n MV_{n+1}M' U_n \\
	&=& \begin{bmatrix} I_{p_1} &  {V_{n+1}^{(1)}}^{-1/2} Y_{n+1}'{V_{n+1}^{(2)}}^{-1/2} \\ {V_{n+1}^{(2)}}^{-1/2} Y_{n+1}{V_{n+1}^{(1)}}^{-1/2} & I_{p_2} \end{bmatrix}.
	\end{eqnarray*}
	\begin{propo} \label{consistencyaux1}
		On the event $\event{E}$, we have
		\begin{eqnarray} \label{consistencyeq1}
		\eigmin{E_n} \geq \rho_0 .
		\end{eqnarray}
	\end{propo}
	
	Then, letting $m=\left\lceil \frac{n}{3} \right\rceil $, define
	\begin{eqnarray*}
	\Sigma_n = V_{m+1}^{(2)}(m) + \sum\limits_{t=m+1}^{n} \trans{2}^{t-m} x^{(2)}(m)x^{(2)}(m)' {\trans{2}'}^{t-m}.
	\end{eqnarray*}
	
	\begin{propo} \label{consistencyaux2}
		For $\tilde{U}_n = \begin{bmatrix} n^{-1/2} I_{p_1} & 0_{p_1 \times p_2} \\ 0_{p_2 \times p_1} & \Sigma_n^{-1/2} \end{bmatrix}$, we have
	\begin{eqnarray} \label{consistencyeq3}
	\Mnorm{\tilde{U}_n^{-1}U_n}{2}^2 \leq \frac{2}{\eigmin{K_1}} + \frac{3}{2} .
	\end{eqnarray}
	\end{propo}
To proceed, define the following matrices:
	\begin{eqnarray*}
		G_n &=& n^{-1} \sum\limits_{t=0}^{n} w(t+1) x^{(1)}(t)' , \\
		H_n &=& n^{-1/2} \sum\limits_{t=0}^{n} w(t+1) x^{(2)}(t)' \Sigma_n^{-1/2}.
	\end{eqnarray*}
		
	\begin{propo} \label{consistencyaux3}
		For matrices $G_n,H_n$ defined above, it holds that
		\begin{eqnarray} 
		\PP{\Mnorm{G_n}{2} > \frac{\epsilon}{\rho_3}} &\leq& \frac{\delta}{2} , \label{consistencyeq4} \\
		\PP{\Mnorm{H_n}{2} > \frac{\epsilon}{\rho_3}} &\leq& \frac{\delta}{2}. \label{consistencyeq8}
		\end{eqnarray}
	\end{propo}
	Finally, since the event $\event{E}$ holds, \eqref{consistencycondition9} implies
	\begin{eqnarray} \label{consistencyeq9}
	\Mnorm{n^{1/2}U_n}{2} \leq 2^{3/2} \eigmin{K_1}^{-1/2}.
	\end{eqnarray}
	This completes the proof as follows. Writing
	\begin{eqnarray*}
		\esttrans{n+1} - \trans{0} &=& \sum\limits_{t=0}^{n} w(t+1)x(t)'V_{n+1}^{-1} \\
		&=& \left[G_n,H_n\right] \left(\tilde{U}_n^{-1} U_n\right) E_n^{-1} n^{1/2} U_n M,
	\end{eqnarray*}
	according to inequalities \eqref{consistencyeq1}, \eqref{consistencyeq3}, \eqref{consistencyeq4}, \eqref{consistencyeq8}, and \eqref{consistencyeq9}, on the event $\event{E}$, with probability at least $1-\delta$,
	\begin{eqnarray*}
		&& \Mnorm{\esttrans{n+1} - \trans{0}}{2} \\
		&\leq& \left(\Mnorm{G_n}{2}+\Mnorm{H_n}{2}\right) \Mnorm{\tilde{U}_n^{-1} U_n}{2} \Mnorm{E_n^{-1}}{2} \Mnorm{n^{1/2}U_n}{2} \Mnorm{M}{2} \\
		&\leq& \frac{2 \epsilon}{\rho_3} \left(\frac{2}{\eigmin{K_1}} + \frac{3}{2}\right)^{1/2} \rho_0^{-1} 2^{3/2} \eigmin{K_1}^{-1/2} \Mnorm{M}{2} \\
		&=& \epsilon,
	\end{eqnarray*}
	which is the desired result.
\endproof

\newpage                                         

\section{Proofs of Auxiliary Results}
\subsection{\bf Proof of Proposition \ref{minpoly}}
Assume $\trans{0}$ is regular. Clearly, the infimum in the definition of $\minpoly{\trans{0}}$ can be taken over $\norm{a}{1}=1$. Further, we will show that there is no polynomial $f$ of degree at most $p-1$, such that $f\left(\trans{0}^{-1}\right)=0$. Note that this finishes the proof as follows. Let
\begin{eqnarray*}
	S_1^p= \{a \in \R^p : \norm{a}{1}=1\}.
\end{eqnarray*}
The function $\mathcal{G}: \R^p \to \R$, defined as $\mathcal{G}(a)=\mincoor{\sum\limits_{i=0}^{p-1}a_{i+1}\Lambda^{-i}}$ is continuous. Since $S_1^p$ is a closed subset of $\R^p$, $\mathcal{G}\left(S_1^p\right) \subset \R$ is closed as well. Therefore, if for all $a \in S_1^p$, we have $\mathcal{G}(a) >0$, then $\inf \mathcal{G}\left(S_1^p\right) >0$, which means $\minpoly{\trans{0}}>0$.

If there is a polynomial $f$, such that $f\left(\trans{0}^{-1}\right)=0$, let $\trans{0}^{-1}=P^{-1} \Gamma P$ be the Jordan decomposition of $\trans{0}^{-1}$, where $\Gamma=\diag{\Gamma_1, \cdots, \Gamma_k}$, and $\Gamma_i$ is a size $m_i$ Jordan matrix of $\gamma_i$, as defined in Definition \ref{Jordandeff}. Now, $f\left(\trans{0}^{-1}\right)=0$ implies $f\left(\Gamma\right)=0$, which in turn yields $f\left(\Gamma_i\right)=0$, for all $i=1,\cdots, k$. As shown in the proof of Proposition \ref{statenorm}, diagonal coordinates of $f\left(\Gamma_i\right)$ are all $f\left(\gamma_i\right)$, i.e. $f\left(\gamma_i\right)=0$. 

Let $f(x)=g(x)\left(x-\gamma_1\right)^{n_1} \cdots \left(x-\gamma_k\right)^{n_k}$, where none of $\gamma_1, \cdots, \gamma_k$ is a root of $g(x)$. We show that for all $i$, $n_i \geq m_i$, so,
\begin{eqnarray*}
	\deg f \geq \sum\limits_{i=1}^{k} n_i \geq \sum\limits_{i=1}^{k} m_i = p,
\end{eqnarray*}
which is a contradiction. Note that by regularity of $\trans{0}$, $\gamma_1, \cdots, \gamma_k$ are distinct, i.e. for $i \neq j$, $\Gamma_i - \gamma_j I_{m_i}$ is invertible (since it is a Jordan matrix of $\gamma_i-\gamma_j \neq 0$). Hence, $f\left(\Gamma_i\right)=0$ implies $\left(\Gamma_i-\gamma_i I_{m_i}\right)^{n_i}=0$. But, as shown in the proof of Proposition \ref{statenorm}, an exponent of size $m$ Jordan matrix of 0 is zero matrix, only if the exponent is not smaller than $m$, i.e. $n_i \geq m_i$, which is the desired result.

Conversely, assume $\trans{0}$ is not regular, i.e. there are $1 \leq i,j \leq k$, such that $\gamma_i=\gamma_j$, and $m_i \geq m_j \geq 1$. Letting $g(x)=\det \left(\trans{0}-xI_p\right)$, define $$f(x)= \frac{1}{x-\gamma_i} g(x),$$ if $\gamma_i$ is real, and $$f(x)= \frac{1}{\left(x-\gamma_i\right)\left(x-\bar{\gamma_i}\right)} g(x),$$ otherwise, where $\bar{\gamma_i}$ is the complex conjugate of $\gamma_i$. 

Clearly, $\deg f \leq p-1$, but we show that $f\left(\trans{0}^{-1}\right)=0$, which leads to $\minpoly{\trans{0}}=0$. Note that the polynomial $f(x)$ can not be a trivial one. As seen in the first part of the proof, it suffices to show that $f\left(\Gamma_\ell\right)=0$, for all $\ell=1,\cdots, k$. If $\ell \neq i,j$, we have $g\left(\Gamma_\ell\right)=0$, so, $f\left(\Gamma_\ell\right)=0$. Since the multiplicity of the root $\gamma_i$ in $g(x)$ is $m_i+m_j$, its multiplicity in $f(x)$ is at least $m_i+m_j-1 \geq m_i$, which is greater than or equal to the dimension of $\Gamma_i$ and $\Gamma_j$. Therefore, $f\left(\Gamma_\ell\right)=0$, for $\ell=i,j$, which completes the proof.
\endproof
\subsection{\bf Proof of Proposition \ref{PDcore}}
We use the following Proposition \cite{lai1983note}.
\begin{propo}
	Let $\{\zeta_n\}_{n=1}^\infty$ be a martingale difference sequence of random variables with respect to the filtration $\{\mathcal{F}_n\}_{n=1}^\infty$, such that
	\begin{eqnarray*}
		\liminf\limits_{n \to \infty} \E{\zeta_n^2 | \mathcal{F}_{n-1}} >0.
	\end{eqnarray*}
	If the real sequence $\{a_n\}_{n=1}^\infty$, satisfies  $\sum\limits_{n=1}^{\infty} a_n^2 \leq \infty$ and $a_n \neq 0$ infinitely often, then $\sum\limits_{n=1}^{\infty} a_n \zeta_n$ has a continuous distribution.
\end{propo}
For an arbitrary row $P_i'$ of $P$, let $v$ be one of the real vectors $\Re \left(P_i\right)$ or $\Im \left(P_i\right)$. Note that since $P$ is invertible, $P_i \neq 0$, and we can assume $v \neq 0$. Taking
\begin{eqnarray*}
	a_n &=& \norm{{\trans{0}'}^{-np}v}{2} , \\
	\zeta_n &=& \frac{1}{a_n} \sum\limits_{i=np-p+1}^{np} v'\trans{0}^{-i}w(i),
\end{eqnarray*}
$a_n \neq 0$ infinitely often, and by $\eigmin{\trans{0}}>1$ we have $\sum\limits_{n=1}^{\infty} a_n^2 < \infty$. Furthermore, by reachability we have
\begin{eqnarray*}
	&& \E{\zeta_n^2} = \frac{1}{a_n^2} \sum\limits_{i=np-p+1}^{np} v'\trans{0}^{-i}C{\trans{0}'}^{-i}v \\
	&=& \frac{1}{\norm{{\trans{0}'}^{-np}v}{2}^2} v'\trans{0}^{-np} \left(\sum\limits_{i=0}^{p-1} \trans{0}^{i}C{\trans{0}'}^{i}\right){\trans{0}'}^{-np}v \\
	&\geq& \eigmin{K(C)}>0.
\end{eqnarray*}

Hence, $v'z(\infty)=v'x(0)+\sum\limits_{n=1}^{\infty} a_n \zeta_n$ has a continuous distribution. Letting $\mathbb{F}_i$ be the Cumulative Distribution Function (CDF) of $\left|P_i' z(\infty)\right|$, $\mathbb{F}_i$ is continuous, and because of $\left|P_i' z(\infty)\right| \geq \left|v' z(\infty)\right|$, one has $\mathbb{F}_i^{-1} \left(\frac{\delta}{p}\right) >0$. Since,
\begin{eqnarray} \label{PDcoreproofeq1}
\PP{\left|P_i'z(\infty)\right| < \mathbb{F}_i^{-1} \left(\frac{\delta}{p}\right)} = \frac{\delta}{p},
\end{eqnarray}
we have
\begin{eqnarray*}
	&& \PP{ \min\limits_{1 \leq i \leq p} \left|P_i'z(\infty)\right| < \min\limits_{1 \leq i \leq p} \mathbb{F}_i^{-1} \left(\frac{\delta}{p}\right) } \\
	&\leq& \sum\limits_{i=1}^{p} \PP{ \left|P_i'z(\infty)\right| <  \mathbb{F}_i^{-1} \left(\frac{\delta}{p}\right) } = \delta,
\end{eqnarray*}
i.e. $\innerproductmin{\trans{0}}{\delta} \geq \min\limits_{1 \leq i \leq p} \mathbb{F}_i^{-1} \left(\frac{\delta}{p}\right) > 0$. 

To proceed, we use the following fact. For two independent random variables $X,Y$, if $X$ has bounded pdf $f_X$, then $X+Y$ has bounded pdf $f_{X+Y}$, and $\sup\limits_{y \in \R}f_{X+Y}(y) \leq \sup\limits_{y \in \R}f_X(y)$. To see this, note that for all $y \in \R$,
\begin{eqnarray*}
	&& f_{X+Y}(y) = \int\limits_{\R} f_X(y-\tau) d\mathbb{P}_Y(\tau) \\ &\leq& \left(\sup\limits_{\tau \in \R}f_X(\tau)\right) \int\limits_{\R}  d\mathbb{P}_Y(\tau) = \sup\limits_{\tau \in \R}f_X(\tau).
\end{eqnarray*}

Now, suppose that the supports of $w(i-p+1), \cdots, w(i)$ are certain subspaces of $\R^p$, and they have bounded pdfs. Then, all of the random variables $v' \trans{0}^{-i+p-1} w(i-p+1), \cdots, v' \trans{0}^{-i}w(i)$ cannot be degenerate. Since otherwise, $\var \left(v' \trans{0}^{-i+j} w(i-j)\right)=0$, for all $j=0, \cdots, p-1$, i.e.
\begin{eqnarray*}
	0 &=& \var \left(v' \sum\limits_{j=0}^{p-1} \trans{0}^{-i+j}w(i-j)\right)= v'\trans{0}^{-i}K(C) {\trans{0}'}^{-i}v \\
	&\geq& \eigmin{K(C)} \norm{{\trans{0}'}^{-i}v}{2}^2 >0,
\end{eqnarray*}
which is a contradiction. Therefore, there exists $j$, such that $\trans{0}^{-i+j}w(i-j)$ lives in a subspace not orthogonal to $v$, i.e. $v'\trans{0}^{-i+j}w(i-j)$ is a continuous random variable, with a bounded pdf (since pdf of $w(i-j)$ is bounded).

Using the aforementioned fact, pdf of $v'z(\infty)$, as well as pdf of $\left|P_iz(\infty)\right|$ which is denoted by $f_i$, are bounded. Letting $\innerproductminconstant{\trans{0}}^{-1} = p \max\limits_{1 \leq i \leq p} \sup\limits_{y \in \R} f_i(y) < \infty$,
\begin{eqnarray*}
	\mathbb{F}_i\left(\innerproductminconstant{\trans{0}}\delta\right) = \int\limits_{0}^{\innerproductminconstant{\trans{0}}\delta} f_i(y) dy \leq  \innerproductminconstant{\trans{0}} \delta \sup\limits_{y \in \R} f_i(y) \leq \frac{\delta}{p},
\end{eqnarray*}
i.e. $\mathbb{F}_i^{-1} \left(\frac{\delta}{p}\right) \geq \innerproductminconstant{\trans{0}} \delta$. 

For normal case, $v' \sum\limits_{j=0}^{p-1} D^{-i+j}w(i-j)$ is normal with pdf $\tilde{f}$, and 
\begin{eqnarray*}
	\var \left(v' \sum\limits_{j=0}^{p-1} \trans{0}^{-i+j}w(i-j)\right)= v'\trans{0}^{-i}K(C) {\trans{0}'}^{-i}v \\
	\geq \eigmin{K(C)} \norm{{\trans{0}'}^{-i}v}{2}^2 >0,
\end{eqnarray*}
i.e.
\begin{eqnarray*}
	\sup\limits_{y \in \R}\tilde{f}(y) \leq \left(2 \pi \eigmin{K(C)}\right)^{-1/2} \norm{{\trans{0}'}^{-i}v}{2}^{-1} \\
	\leq \left(\frac{\eigmax{{\trans{0}}^{i}{\trans{0}'}^{i}}}{2 \pi \eigmin{K(C)}}\right)^{1/2} \norm{v}{2}^{-1}.
\end{eqnarray*}
Denote the right hand side of the above by $\frac{1}{2bp}$. By the fact mentioned before, $v'z(\infty)$ has a pdf, denoted by $f$, which is bounded by $\frac{1}{2bp}$. Letting $\mathbb{F}$ be CDF of $\left|v' z(\infty)\right|$, we have
\begin{eqnarray*}
	\mathbb{F}\left(b \delta \right) = \int\limits_{-b\delta}^{b \delta} f(y)dy \leq 2 b \delta \sup\limits_{-b\delta \leq y \leq b \delta}f(y) \\ \leq 2 b \delta \sup\limits_{y \in \R}f_1(y) \leq \frac{\delta}{p},
\end{eqnarray*}
which by $\left|P_i'z(\infty)\right| \geq \left|v'z(\infty)\right|$, implies $\mathbb{F}_i^{-1}\left(\frac{\delta}{p}\right) \geq b \delta$. Plugging in \eqref{PDcoreproofeq1}, we get the desired result.
\endproof

\subsection{\bf Proof of Fact \ref{unitrootexclusion}}
Assume $X \in \R^{p \times p}$ has an eigenvalue of unit size, denoted by $\lambda \in \C, \left| \lambda \right|=1$. Further, define the space of eigenvectors in $\C^p$ as follows. First, consider the equivalence relation $\sim$ on $\C^p$, defined as $$x \sim y \text{, if } x=cy \text{ for some } c \in \C, c \neq 0.$$  
Letting $S= \frac{\C^p }{\sim}$ be the direction space in $\C^p$, we have $\dimension{\C}{S}=p-1$, i.e. $\dimension{\R}{S}=2p-2.$
Note that for every matrix $Y \in \C^{p \times p}$ and every vector $v \in \C^p$, $Yv=0$ if and only if $Y \tilde{v}=0$ for every $\tilde{v} \sim v$. Thus, $\det \left( X - \lambda I_p \right)=0$ implies that there is $v \in S, v \neq 0$, such that
\begin{eqnarray} \label{2unitrootexclusioneq1}
Xv = \lambda v
\end{eqnarray}
Denote the set of all matrices $X$ satisfying \eqref{2unitrootexclusioneq1} by $\mathcal{X} \left(\lambda, v \right) \subset \R^{p \times p}$. Separating real and imaginary parts, we get
\begin{eqnarray*}
	X \Re\left(v\right) &=& \Re\left( \lambda v\right), \\
	X \Im\left(v\right) &=& \Im \left( \lambda v\right).
\end{eqnarray*}
Then, we partition $S$ to $S=S_1 \cup S_2, S_1 \cap S_2 = \emptyset,$ where
\begin{eqnarray*}
	S_1 &=& \{ v \in S : \Re\left(v\right), \Im\left(v\right) \text{ are in-line } \}, \\
	S_2 &=& \{ v \in S : \Re\left(v\right), \Im\left(v\right) \text{ are not in-line } \}. 
\end{eqnarray*}
Whenever $v \in S_2$, for $j=1,\cdots,p$, the $j$-th row of $X$ needs to be in the intersection of two nonparallel hyperplanes $\mathcal{P}_1,\mathcal{P}_2 \subset \R^p$, where
\begin{eqnarray*}
	\mathcal{P}_1 &=& \left\{ y \in \R^p: y'\Re(v)=\Re(\lambda v)_j \right\}, \\
	\mathcal{P}_2 &=& \left\{ y \in \R^p: y'\Im(v)=\Im(\lambda v)_j \right\}.
\end{eqnarray*}
Since $\dimension{\R}{\mathcal{P}_1} \leq p-1$, $\dimension{\R}{\mathcal{P}_2} \leq p-1$, and $v \in S_2$ we have $ \dimension{\R}{\mathcal{P}_1 \cap \mathcal{P}_2} \leq p-2.$
Therefore, for $v \in S_2$, we have $\dimension{\R}{\mathcal{X} \left(\lambda, v \right)} \leq p(p-2)$. Because of $\dimension{\R}{\left|\lambda\right|=1}=1$, and $\dimension{\R}{S_2} \leq 2p-2$, we have 
\begin{eqnarray} \label{2unitrootexclusioneq2}
\dimension{\R}{\bigcup\limits_{\left|\lambda\right|=1, v \in S_2} \mathcal{X} \left(\lambda,v\right)} \leq p^2-1 .
\end{eqnarray}

On the other hand, for $v \in S_1$, there is a real number, say $\alpha(v)$, such that $\Im \left(v\right) = \alpha(v) \Re \left(v\right)$. So, $\dimension{\R}{S_1}=p-1$, and for $v \in S_1$, we have $\mathcal{P}_1=\mathcal{P}_2$, i.e. $\dimension{\R}{\mathcal{X} \left(\lambda, v \right)} \leq p(p-1),$
and
\begin{eqnarray*}
	0 = \alpha(v) X \Re\left(v\right) - X \Im\left(v\right) = \alpha(v) \Re\left( \lambda v\right) - \Im\left(\lambda v\right) \\
	= \left(1+\alpha(v)^2\right) \Im \left(\lambda\right) \Re\left(v\right),
\end{eqnarray*}
i.e. either $\Im \left(\lambda\right)=0$, or $\Re\left(v\right)=0$. Note that the latter case is impossible because it implies $v=0$. So, since $\left\{ \left|\lambda\right|=1, \Im \left(\lambda\right)=0 \right\} = \{1,-1\}$ is of dimension zero, 
\begin{eqnarray} \label{2unitrootexclusioneq3}
\dimension{\R}{\bigcup\limits_{\lambda=-1,1} \mathcal{X} \left(\lambda,v\right)} \leq p^2-1 .
\end{eqnarray}
Therefore, letting $\mathcal{X}= \bigcup\limits_{\lambda, v} \mathcal{X} \left(\lambda,v\right)$, \eqref{2unitrootexclusioneq2} and \eqref{2unitrootexclusioneq3} imply $\dimension{\R}{\mathcal{X}} \leq p^2-1$, i.e. $\mathcal{X}$ is of zero Lebesgue measure in $\R^{p \times p}$.

To prove that irregular matrices are of zero Lebesgue measure, for $\left| \lambda \right|>1$ define
\begin{eqnarray*}
	\mathcal{Y} \left( \lambda \right) = \left\{ Y \in \R^{p \times p}: \rank{Y - \lambda I_p} < p-1 \right\}.
\end{eqnarray*}
First we show that for a fixed matrix $Y=\left[Y_1,\cdots, Y_p\right]$, there are at most $p-1$ values of $\lambda$ such that $Y \in \mathcal{Y} \left(\lambda\right)$. Let $e_1, \cdots, e_p$ be the standard basis of $\R^p$. If $Y \in \mathcal{Y} \left(\lambda_0 \right)$, two of $Y_i- \lambda_0 e_i, i=1,\cdots,p$, such as $Y_{p-1}-\lambda_0 e_{p-1}, Y_p- \lambda_0 e_p$, can be written as a linear combinations of the others. There are at most $p-1$ values of $\lambda_0$ for which $Y_{p-1}- \lambda_0 e_{p-1}$ is a linear combination of $ Y_1- \lambda_0 e_1, \cdots, Y_{p-2}- \lambda_0 e_{p-2}$, since for every such a $\lambda_0$, $\det \left(\tilde{Y}\right)=0$, where $\tilde{Y}$ is the square matrix whose columns are $ Y_1- \lambda_0 e_1, \cdots, Y_{p-1}- \lambda_0 e_{p-1},$
removing an arbitrary row. Note that $\det \left(\tilde{Y}\right)$ is a polynomial of degree $p-1$. 

Now, denote those values of $\lambda$ by $\lambda_1 \left( Y \right), \cdots, \lambda_{m} \left( Y \right)$, where $m \leq p-1$. For every $i=1,\cdots,m$, the dimension of subspace $\mathcal{P}_i$ spanned by $ Y_1- \lambda_i \left(Y\right)e_1, \cdots, Y_{p-1}-\lambda_i \left(Y\right)e_{p-1}, e_p$ 
is at most $p-1$, which leads to
\begin{eqnarray*} 
	\dimension{\R}{\bigcup\limits_{i=1}^m \mathcal{P}_i} \leq p-1.
\end{eqnarray*}
Because $\lambda_i \left(Y\right)$ is uniquely determined by $Y_1, \cdots, Y_{p-1}$, so is $\mathcal{P}_i$. Therefore, 
\begin{eqnarray*}
	&& \dimension{\R}{\bigcup\limits_{\lambda} \mathcal{Y} \left(\lambda\right)} \\
	&\leq& \dimension{\R}{\left[Y_1, \cdots, Y_{p-1}\right]} + \dimension{\R}{\bigcup\limits_{i=1}^m \mathcal{P}_i} \\ 
	&\leq& p (p-1) + p-1 = p^2-1,
\end{eqnarray*}
which completes the proof. 
\endproof

\subsection{\bf Proof of Proposition \ref{noisebound}}
First, note that for all $y>0 ; i=1,\cdots,p ; t=1,\cdots, n$, by Assumption \ref{tailcondition} we have
\begin{eqnarray*}
	\PP{\left|w_i(t)\right| > \noisemax{n}{\delta}} &\leq& \tailconst \exp \left(-\frac{\noisemax{n}{\delta}^\tailexp}{\tailcoeff}\right) \\
	&=& \tailconst \exp \left(-\frac{\tailcoeff \log \frac{ \tailconst np}{\delta}}{\tailcoeff}\right) = \frac{\delta}{np} . 
\end{eqnarray*}
Using a union bound, we get
\begin{eqnarray*}
	\PP{\event{W}^c} \leq \sum\limits_{t=1}^{n} \sum\limits_{i=1}^{p} \PP{\left|w_i(t)\right| > \noisemax{n}{\delta}} \leq \delta.
\end{eqnarray*}
\endproof

\subsection{\bf Proof of Proposition \ref{statenorm}}	
First, let $\trans{0}=P^{-1}\Lambda P$ be its Jordan decomposition. The behavior of $\Mnorm{\Lambda}{\infty}$ is determined by the blocks of $\Lambda$. In fact, letting $\Lambda=\diag{\Lambda_1, \cdots, \Lambda_k}$, the definition of $\Mnorm{\cdot}{\infty}$ implies $\Mnorm{\Lambda}{\infty} \leq \max \limits_{1 \leq i \leq k} \Mnorm{\Lambda_i}{\infty}$. Then, to control the norm of an exponent of an arbitrary block, we show that $\Mnorm{\Lambda_i^t}{\infty} \leq \MJordanconst{t}{\Lambda_i}$. For this purpose, note that for $k=0,1, \cdots$,
\begin{eqnarray*}
	\Lambda_i^k = \begin{bmatrix}
		\lambda_i^k & {k \choose 1} \lambda_i^{k-1}  & \cdots & {k \choose m-1} \lambda_i^{k-m+1} \\
		0 & \lambda_i^k & \cdots & {k \choose m-2} \lambda_i^{k-m+2} \\
		\vdots & \vdots & \vdots & \vdots \\
		0 & \cdots & 0 & \lambda_i^k
	\end{bmatrix},
\end{eqnarray*}
and let $v \in \C^{m_i}$ be such that $\norm{v}{\infty}=1$. For $\ell=1,\cdots,m_i$, the $\ell$-th coordinate of $\Lambda_i^t v$ is $\sum\limits_{j=0}^{m_i-\ell} {t \choose j} \lambda_i^{t-j}v_{j+\ell+1}$, which, because of ${t \choose j} \leq \frac{t^j}{j!}$, is at most $\MJordanconst{t}{\Lambda_i}$. Therefore, because of $\Lambda^t=\diag{\Lambda_1^t, \cdots, \Lambda_k^t}$, we have $\Mnorm{\Lambda^t}{\infty} \leq \MJordanconst{t}{\Lambda}$. Now, by $x(t)=\trans{0}^t x(0)+ \sum\limits_{i=1}^{t}\trans{0}^{t-i}w(i)$, on the event $\event{W}$ we have 
\begin{eqnarray*}
	&& \norm{x(t)}{2} \\
	&=& \norm{P^{-1}\Lambda^tP x(0)+ \sum\limits_{i=1}^{t}P^{-1}\Lambda^{t-i}Pw(i)}{2} \\
	&\leq& \Opnorm{P^{-1}}{\infty}{2} \left(\Mnorm{\Lambda^t}{\infty} \norm{Px(0)}{\infty}+ \sum\limits_{i=1}^{t}\norm{\Lambda^{t-i}Pw(i)}{\infty}\right) \\
	&\leq& \MJordanconst{}{\trans{0}} \left(\norm{x(0)}{\infty} + \noisemax{n}{\delta} \right).
\end{eqnarray*}
\endproof

\subsection{\bf Proof of Proposition \ref{empiricalcov}}
In this proof, we use the following Matrix Bernstein inequality \cite{tropp2012user}:
\begin{propo} \label{MBernstein}
	Let $X_i \in \R^{p \times p}, i=1,\cdots, n$ be a sequence of independent symmetric random matrices. Assume for all $i=1,\cdots, n$, we have $\E{X_i}=0$ and $\eigmax{X_i} \leq \ssconstant$. Then, for all $y \geq 0$ we have
	\begin{eqnarray*}
		\PP{\eigmax{\sum\limits_{i=1}^{n}X_i} \geq y} \leq 2p \!\ \exp \left(-\frac{3y^2}{6\sigma^2 + 2 \ssconstant y}\right),
	\end{eqnarray*}
	where $\sigma^2 = \eigmax{\sum\limits_{i=1}^{n}\E{X_i^2}}$.
\end{propo}
Letting $X_i=w(i)w(i)'-C$, and $\ssconstant = p \noisemax{n}{\delta}^2$, clearly $\E{X_i}=0$, and
\begin{eqnarray*}
	\sigma^2 &=& \eigmax{\sum\limits_{i=1}^{n}\E{X_i^2}} \\
	&\leq& \sum\limits_{i=1}^{n}\eigmax{\E{\norm{w(i)}{2}^2 w(i)w(i)'}-C^2} \\
	&\leq& n \ssconstant \eigmax{C}. 
\end{eqnarray*}
On $\event{W}$, we have $\max\limits_{1 \leq i \leq n}\norm{w(i)}{2}^2 \leq \ssconstant$. Therefore, \eqref{empiricalcoveq1} implies
\begin{eqnarray*}
	&& \PP{\eigmax{C_n-C}>\epsilon} = \PP{\eigmax{\sum\limits_{i=1}^{n}X_i}>n\epsilon} \\
	&\leq& 2p \!\ \exp \left(- \frac{3n\epsilon^2}{6 \ssconstant \eigmax{C}+2 \ssconstant \epsilon}\right) \leq \delta.
\end{eqnarray*}
\endproof

\subsection{\bf Proof of Proposition \ref{crossproduct}}
In this proof, we use the following Matrix Azuma inequality \cite{tropp2012user}:
\begin{propo} \label{MAzuma}
	Let $X_i \in \R^{p \times p}, i=1,\cdots, n$ be a martingale difference sequence of symmetric matrices, i.e. for some filtration  $\{\mathcal{F}_i\}_{i=0}^n$, $X_i$ is $\mathcal{F}_i$-measurable and $\E{X_{i+1} | \mathcal{F}_i}=0$. Assume for fixed symmetric matrices $M_i, i=1,\cdots, n$, all matrices $M_i^2-X_i^2$ are positive semidefinite. Then, for all $y \geq 0$ we have
	\begin{eqnarray*}
		\PP{\eigmax{\sum\limits_{i=1}^{n}X_i} \geq y} \leq 2p \!\ \exp \left(-\frac{y^2}{8\sigma^2}\right),
	\end{eqnarray*}
	where $\sigma^2 = \eigmax{\sum\limits_{i=1}^{n}M_i^2}$.
\end{propo}
Letting $X_i = \trans{0}x(i-1)w(i)'+w(i)x(i-1)'\trans{0}',
\mathcal{F}_i = \sigma \left(w(1), \cdots,w(i)\right),
M_i = 2 p^{1/2} \predbound{n}{\delta} I_p ,$
clearly, $\E{X_{i+1}|\mathcal{F}_i}=0$, and $M_i^2-X_i^2$ is positive semidefinite, since by Propositions \ref{noisebound}, \ref{statenorm}, on $\event{W}$ we have 
\begin{eqnarray*}
	\max\limits_{1 \leq i \leq n} \norm{w(i)}{2} &\leq& p^{1/2} \noisemax{n}{\delta} , \\
	\max\limits_{0 \leq i \leq n-1} \norm{x(i)}{2} &\leq& \MJordanconst{}{\trans{0}} \left(\norm{x(0)}{\infty} +\noisemax{n}{\delta} \right).
\end{eqnarray*}
Therefore, $\sigma^2 = 4np \predbound{n}{\delta}^2$, and 
\begin{eqnarray*}
	&& \PP{\eigmax{U_n}>\epsilon} = \PP{\eigmax{\sum\limits_{i=1}^{n}X_i}>n\epsilon} \\
	&\leq& 2p \!\ \exp \left(- \frac{n\epsilon^2}{32p \predbound{n}{\delta}^2}\right) \leq \delta.
\end{eqnarray*}
\endproof

\subsection{\bf Proof of Proposition \ref{zinfinitynormbound}}
First, according to Proposition \ref{noisebound},
\begin{eqnarray*}
	&& \PP{\norm{w(t)}{\infty} \leq \noisemax{1}{\frac{\delta}{2t^2}}, \forall t=1,2,\cdots} \\
	&=& 1- \PP{\norm{w(t)}{\infty} > \noisemax{1}{\frac{\delta}{2t^2}}, \exists t=1,2,\cdots} \\
	&\geq& 1-\sum\limits_{t=1}^{\infty} \PP{\norm{w(t)}{\infty} > \noisemax{1}{\frac{\delta}{2t^2}}} \\
	&\geq& 1- \sum\limits_{t=1}^{\infty} \frac{\delta}{2t^2} \\
	&>& 1- \delta.
\end{eqnarray*}
Then, similar to the proof of Proposition \ref{statenorm}, we have $\Mnorm{\Lambda^{-t}}{\infty} \leq \MJordanconst{t}{\Lambda^{-1}}$, i.e. for all $n=1,2,\cdots$,  
\begin{eqnarray*}
	&& \norm{z(n)}{2} \leq \sum\limits_{t=1}^{\infty} \Opnorm{\trans{0}^{-t}}{\infty}{2} \norm{w(t)}{\infty} \leq \norm{x(0)}{2} \\
	&+& \Opnorm{P^{-1}}{\infty}{2} \Mnorm{P}{\infty} \sum\limits_{t=1}^{\infty} \MJordanconst{t}{\Lambda^{-1}} \noisemax{1}{\frac{\delta}{2t^2}} = \zinfinitybound{\trans{0}}{\delta},
\end{eqnarray*}
with probability at least $1-\delta$.
\endproof

\subsection{\bf Proof of Proposition \ref{explosiveminaux1}}
On the event $\event{W}$, similar to the proof of Proposition \ref{statenorm}, for all $t=1,\cdots, n$ we have
\begin{eqnarray*} 
	&& \norm{z(n)-z(n-t)}{2} \leq  \sum\limits_{i=n-t+1}^{n}\norm{\trans{0}^{-i}w(i)}{2} \\
	&\leq& \Opnorm{P^{-1}}{\infty}{2} \Mnorm{P}{\infty} \noisemax{n}{\delta} \sum\limits_{i=n-t+1}^{n} \MJordanconst{i}{\Lambda^{-1}}. 
\end{eqnarray*}
Similarly, noting that $\MJordanconst{t}{\Lambda'^{-1}}=\MJordanconst{t}{\Lambda^{-1}}$, for $t=0, 1,2,\cdots$, we get
\begin{eqnarray} \label{explosiveminproofeq0} 
\Mnorm{\trans{0}'^{-t}}{2} \leq \Opnorm{P'}{\infty}{2} \Mnorm{{P'}^{-1}}{\infty} \MJordanconst{t}{\Lambda^{-1}}, 
\end{eqnarray}
Thus, using \eqref{Jordaneq1},
\begin{eqnarray*}
	& & \sum\limits_{t=0}^{n} \norm{z(n-t)-z(n)}{2} \Mnorm{{\trans{0}'}^{-t}}{2}^2 \\
	&\leq& \sum\limits_{t=0}^{n/3} \norm{z(n-t)-z(n)}{2} \Mnorm{{\trans{0}'}^{-t}}{2}^2 \\
	&+& \sum\limits_{t=n/3}^{n} \norm{z(n-t)-z(n)}{2} \Mnorm{{\trans{0}'}^{-t}}{2}^2 \\
	&\leq& \Opnorm{P^{-1}}{\infty}{2} \Mnorm{P}{\infty} \noisemax{n}{\delta} \left(\sum\limits_{i=2n/3}^{n} \MJordanconst{i}{\Lambda^{-1}}\right) \sum\limits_{t=0}^{n/3}  \Mnorm{{\trans{0}'}^{-t}}{2}^2 \\
	&+& \sum\limits_{t=n/3}^{n} \norm{z(n-t)-z(n)}{2} \left(\Opnorm{P'}{\infty}{2} \Mnorm{{P'}^{-1}}{\infty} \MJordanconst{t}{\Lambda^{-1}}\right)^2 \\
	&\leq& \frac{1}{2} \rho_1 \noisemax{n}{\delta} n^{2\mult{\trans{0}}} \eigmin{\trans{0}}^{-2n/3} , 
\end{eqnarray*}
which by \eqref{explosivemincondition1} implies
\begin{eqnarray} \label{explosiveminproofeq1}
\sum\limits_{t=0}^{n} \norm{z(n-t)-z(n)}{2} \Mnorm{{\trans{0}'}^{-t}}{2}^2 \leq \frac{\epsilon}{2\zinfinitybound{\trans{0}}{\delta}}.
\end{eqnarray}
By $x(t)=\trans{0}^{t}z(t)$, since
\begin{eqnarray*}
	&& U_n - F_n \\
	&=& \sum\limits_{t=0}^{n} \trans{0}^{-n} x(t)x(t)' {\trans{0}'}^{-n} - \trans{0}^{-t} z(n)z(n)' {\trans{0}'}^{-t} \\
	&=& \sum\limits_{t=0}^{n} \trans{0}^{-n+t} z(t)z(t)' {\trans{0}'}^{-n+t} - \trans{0}^{-t} z(n)z(n)' {\trans{0}'}^{-t} \\
	&=& \sum\limits_{t=0}^{n} \trans{0}^{-t} z(n-t)z(n-t)' {\trans{0}'}^{-t} - \trans{0}^{-t} z(n)z(n)' {\trans{0}'}^{-t} \\
	&=& \sum\limits_{t=0}^{n} \trans{0}^{-t} \left(z(n-t)z(n-t)' - z(n)z(n)'\right) {\trans{0}'}^{-t},
\end{eqnarray*}
it holds that
\begin{eqnarray*}
	&& \eigmax{U_n - F_n} \\
	&\leq& \sum\limits_{t=0}^{n} \norm{z(n-t)-z(n)}{2} \norm{z(n-t)+z(n)}{2} \Mnorm{{\trans{0}'}^{-t}}{2}^2 \\
	&\leq& 2 \left(\sup\limits_{1 \leq n \leq \infty} \norm{z(n)}{2}\right) \sum\limits_{t=0}^{n} \norm{z(n-t)-z(n)}{2} \Mnorm{{\trans{0}'}^{-t}}{2}^2.
\end{eqnarray*}
Using \eqref{explosiveminproofeq1}, and Proposition \ref{zinfinitynormbound}, on the event $\event{W} \cap \event{V}$, we get
\begin{eqnarray*} 
	\eigmax{U_n - F_n} \leq \frac{\epsilon}{2}.
\end{eqnarray*}
\endproof

\subsection{\bf Proof of Proposition \ref{explosiveminaux2}}
One can use the same argument used in the proof of Proposition \ref{zinfinitynormbound}, to show that the following holds with probability at least $1-\delta$. 
\begin{eqnarray*}
	\norm{z(\infty)-z(n)}{2} &\leq& \Mnorm{\trans{0}^{-n}}{2}\norm{\sum\limits_{t=1}^{\infty}\trans{0}^{-t}w(n+t)}{2} \\
	&\leq& \Mnorm{\trans{0}^{-n}}{2} \zinfinitybound{\trans{0}}{\delta}.
\end{eqnarray*}
Therefore, using \eqref{explosiveminproofeq0}, on the event $\event{V}$, with probability at least $1-\delta$ we have
\begin{eqnarray*}
	&&\eigmax{F_\infty - F_n} \\ &\leq& \sum\limits_{t=0}^{n} \norm{z(\infty)-z(n)}{2} \norm{z(\infty)+z(n)}{2} \Mnorm{{\trans{0}'}^{-t}}{2}^2 \\
	&+&  \sum\limits_{t=n+1}^{\infty} \norm{\trans{0}^{-t} z(\infty)}{2}^2 \\
	&\leq& 2 \zinfinitybound{\trans{0}}{\delta} \norm{z(\infty)-z(n)}{2} \sum\limits_{t=0}^{n} \Mnorm{{\trans{0}'}^{-t}}{2}^2 \\
	&+&  \zinfinitybound{\trans{0}}{\delta}^2 \Mnorm{\trans{0}^{-n}}{2}^2 \sum\limits_{t=1}^{\infty} \Mnorm{{\trans{0}'}^{-t}}{2}^2 \\
	&\leq& \MJordanconst{}{\trans{0}'^{-1}}^2 \zinfinitybound{\trans{0}}{\delta}^2 \left(2  + \Mnorm{\trans{0}^{-n}}{2}\right)\Mnorm{\trans{0}^{-n}}{2} \\
	&\leq& \MJordanconst{}{\trans{0}'^{-1}}^2 \zinfinitybound{\trans{0}}{\delta}^2 \left(2 + \MJordanconst{}{\trans{0}^{-1}}\right) \\
	&\times& \Opnorm{P^{-1}}{\infty}{2} \Mnorm{P}{\infty} \MJordanconst{n}{\Lambda^{-1}}.
\end{eqnarray*}
By \eqref{Jordaneq1}, \eqref{explosivemincondition2} implies that on $\event{V}$, with probability at least $1-\delta$,
\begin{eqnarray*} 
	\eigmax{F_\infty - F_n} \leq \frac{\epsilon}{2}.
\end{eqnarray*}
\endproof

\subsection{\bf Proof of Proposition \ref{explosiveminauxiliary}}
If $\minpoly{\trans{0}}=0$, obviously the statement holds. So, assume $\minpoly{\trans{0}}>0$. Letting $\trans{0}=P^{-1}\Lambda P$ be its Jordan decomposition, we have $f\left(\trans{0}^{-1}\right)=P^{-1} f\left(\Lambda^{-1}\right)P$. The matrix $\Lambda$ is block diagonal, thus, $f\left(\Lambda^{-1}\right)$ is block diagonal as well. Further, every block of $\Lambda^{-i}$, as well as every block of $f\left(\Lambda^{-1}\right)$, is upper triangular (see proof of Proposition \ref{statenorm}). Therefore, since $\minpoly{\trans{0}}>0$, there is at least one row of $f\left(\Lambda^{-1}\right)$, which has exactly one nonzero entry. 

This nonzero coordinate, by the definition of $\minpoly{\trans{0}}$, is in magnitude at least $\norm{a}{1} \Opnorm{P}{2}{\infty} \minpoly{\trans{0}}$. On the other hand, by the definition of $\innerproductmin{\trans{0}}{\delta}$, all coordinates of the vector $Pz(\infty)$ are in magnitude at least $\innerproductmin{\trans{0}}{\delta}$, with probability at least $1-\delta$. 

So, with probability at least $1-\delta$, the vector $u=f\left(\Lambda^{-1}\right) P z(\infty)$ has a coordinate, which is in magnitude at least $\norm{a}{1} \Opnorm{P}{2}{\infty} \minpoly{\trans{0}} \innerproductmin{\trans{0}}{\delta}$. This implies the desired inequality, because 
\begin{eqnarray*}
	&& \norm{a}{1} \Opnorm{P}{2}{\infty} \minpoly{\trans{0}} \innerproductmin{\trans{0}}{\delta} \leq \norm{f\left(\Lambda^{-1}\right) P z(\infty)}{\infty} \\
	&=& \norm{P f\left(\trans{0}^{-1}\right) z(\infty)}{\infty} \leq \Opnorm{P}{2}{\infty} \norm{f\left(\trans{0}^{-1}\right) z(\infty)}{2}.
\end{eqnarray*}
\endproof

\subsection{\bf Proof of Proposition \ref{zinfinityfullrankness}}
Let $\trans{0}=P^{-1}\Lambda P$ be the Jordan decomposition of $\trans{0}$. Whenever
$$\rank{\left[z(\infty), \cdots, \trans{0}^{-p+1}z(\infty)\right]}<p,$$ 
there is a nontrivial real polynomial $f$ of degree at most $p-1$, such that $f\left(\trans{0}^{-1}\right)z(\infty)= P^{-1} f\left(\Lambda^{-1}\right) P z(\infty)=0$. Since $\minpoly{\trans{0}}>0$, similar to the proof of Proposition \ref{explosiveminauxiliary}, there is at least one row of $f\left(\Lambda^{-1}\right)$, say the $i$-th row, which has exactly one nonzero coordinate, say the $ij$-th entry. 

Therefore, since the $i$-th coordinate of the vector $f\left(\Lambda^{-1}\right) P z(\infty)=0$ is zero, the $j$-th coordinate of $P z(\infty)=0$ must be zero; i.e. $P_j' z(\infty)=0$, where $P=\left[P_1, \cdots, P_p\right]'$. So, the desired result holds because
\begin{eqnarray*}
	&& \PP{\rank{\left[z(\infty), \cdots, D^{-p+1}z(\infty)\right]}<p}\\
	&=& \PP{\exists j : P_j' z(\infty)=0} =0.
\end{eqnarray*}
To justify the last equality above, note that similar to the proof of Proposition \ref{PDcore}, for all $j=1,\cdots, p$, $\left|P_j'z(\infty)\right|$ has a continuous distribution, which yields $\PP{\left|P_j' z(\infty)\right|=0} =0$.
\endproof

\subsection{\bf Proof of Proposition \ref{newmatricesreachability}}
Assume $\tilde{v} \in \R^{p_1}, \tilde{v} \neq 0$. We show that $\left[\trans{1},C_{11}\right]$ is reachable. Defining $v=\left[\tilde{v}',0_{1 \times p_2}\right]' \in \R^p$,
\begin{eqnarray*}
	0 &<& \norm{M'v}{2}^2 \eigmin{K(C)} \leq v' M K(C) M' v \\
	&=& v' \left(\sum\limits_{j=0}^{p-1} \tilde{A}^j \tilde{C} \tilde{A}'^j\right) v = \tilde{v}' \left(\sum\limits_{j=0}^{p-1} \trans{1}^j C_{11} {\trans{1}'}^j\right) \tilde{v},
\end{eqnarray*}
so, the matrix $\sum\limits_{j=0}^{p-1} \trans{1}^j C_{11} {\trans{1}'}^j$ is positive definite, or equivalently,
\begin{eqnarray} \label{newmatricesreachabilityeq1}
\rank{\left[C_{11}^{1/2}, \trans{1}C_{11}^{1/2}, \cdots, \trans{1}^{p-1}C_{11}^{1/2}\right]} = p_1.
\end{eqnarray}
But, by the Cayley-Hamilton theorem, \eqref{newmatricesreachabilityeq1} is equivalent to
\begin{eqnarray*}
	\rank{\left[C_{11}^{1/2}, \cdots, \trans{1}^{p_1-1}C_{11}^{1/2}\right]} = p_1,
\end{eqnarray*}
which is nothing but the reachability of $\left[\trans{1},C_{11}\right]$. The proof for $\left[\trans{2},C_{22}\right]$ is similar.
\endproof

\subsection{\bf Proof of Proposition \ref{consistencyaux1}}

Let $m=\left\lceil \frac{n}{3} \right\rceil $, and $v_i \in \R^{p_i}$ for $i=1,2$, $v=\begin{bmatrix} v_1 \\ v_2\end{bmatrix} \in \R^p$, $\norm{v}{2}=1$. Then,
\begin{eqnarray*}
	v'E_nv &=& 1 + 2 v_2'{V_{n+1}^{(2)}}^{-1/2} Y_{n+1}{V_{n+1}^{(1)}}^{-1/2} v_1 \\
	&=& 1 + 2 \term{1} + 2 \term{2} , 
\end{eqnarray*}
where
\begin{eqnarray*}
	\term{1} &=& \sum\limits_{t=0}^{m} v_2'{V_{n+1}^{(2)}}^{-1/2} x^{(2)}(t)v_1'{V_{n+1}^{(1)}}^{-1/2} x^{(1)}(t) , \\
	\term{2} &=& \sum\limits_{t=m+1}^{n} v_2'{V_{n+1}^{(2)}}^{-1/2} x^{(2)}(t)v_1'{V_{n+1}^{(1)}}^{-1/2} x^{(1)}(t).
\end{eqnarray*}
By the Cauchy-Schwarz inequality,
\begin{eqnarray*}
	\term{1}^2 &\leq& \left(v_1'{V_{n+1}^{(1)}}^{-1/2} {V_{m+1}^{(1)}} {V_{n+1}^{(1)}}^{-1/2} v_1\right) \\
	&\times&\left(v_2'{V_{n+1}^{(2)}}^{-1/2} {V_{m+1}^{(2)}} {V_{n+1}^{(2)}}^{-1/2} v_2\right) \\
	&\leq& \norm{v_1}{2}^2 \norm{v_2}{2}^2 \\ 
	&\times& \eigmax{{V_{n+1}^{(2)}}^{-1/2}\trans{2}^{n} \trans{2}^{-n} V_{m+1}^{(2)}{\trans{2}'}^{-n} {\trans{2}'}^{n} {V_{n+1}^{(2)}}^{-1/2}} \\
	&\leq& \norm{v_1}{2}^2 \norm{v_2}{2}^2 \eigmax{\trans{2}^{-n} V_{m+1}^{(2)}{\trans{2}'}^{-n}} \\ 
	&\times& \eigmax{{V_{n+1}^{(2)}}^{-1/2}\trans{2}^{n} {\trans{2}'}^{n} {V_{n+1}^{(2)}}^{-1/2}} . 
\end{eqnarray*} 
Letting $z(t)=\trans{2}^{-t}x^{(2)}(t)$, by Proposition \ref{zinfinitynormbound}, we have
\begin{eqnarray*}
	&& \eigmax{\trans{2}^{-n} V_{m+1}^{(2)}{\trans{2}'}^{-n}} \\
	&\leq& \sum\limits_{t=0}^{m} \norm{z(t)}{2}^2 \Mnorm{{\trans{2}'}^{-n+t}}{2}^2 \\
	&\leq& \zinfinitybound{\trans{2}}{\delta}^2 \Opnorm{P'}{\infty}{2}^2 \Mnorm{{P'}^{-1}}{\infty}^2 \sum\limits_{t=0}^{m} \MJordanconst{n-t}{\Lambda_2^{-1}}^2 , \\
	&& \eigmax{{V_{n+1}^{(2)}}^{-1/2}\trans{2}^{n} {\trans{2}'}^{n} {V_{n+1}^{(2)}}^{-1/2}} \\
	&\leq& \tr{{V_{n+1}^{(2)}}^{-1/2}\trans{2}^{n} {\trans{2}'}^{n} {V_{n+1}^{(2)}}^{-1/2}} = \tr{{\trans{2}'}^{n} {V_{n+1}^{(2)}}^{-1}\trans{2}^{n}} \\
	&\leq& p\eigmin{{\trans{2}}^{-n} {V_{n+1}^{(2)}} {\trans{2}'}^{-n}}^{-1} \leq 2p \minpoly{\trans{2}}^{-2} \innerproductmin{\trans{2}}{\delta}^{-2}.
\end{eqnarray*} 
According to \eqref{Jordaneq1}, 
\begin{eqnarray*}
	\sum\limits_{t=0}^{m} \MJordanconst{t}{\Lambda_2^{-1}}^2 \leq e^{2\eigmin{\trans{2}}} n^{2\mult{\trans{2}}-1} \eigmin{\trans{2}}^{2m-2n}.
\end{eqnarray*}
So, by \eqref{consistencycondition3}, we have
\begin{eqnarray} 
\term{1} &\leq& \norm{v_1}{2} \norm{v_2}{2} \rho_1 n^{\mult{\trans{2}}-1/2} \eigmin{\trans{2}}^{-2n/3} \\ 
&\leq& \rho_0 \norm{v_1}{2} \norm{v_2}{2}.\label{eqnarray1}
\end{eqnarray}
Similarly, an application of the Cauchy-Schwarz inequality implies
\begin{eqnarray} 
\term{2} &\leq& \norm{v_2}{2} \left(\sum\limits_{t=m+1}^{n} \left(v_1'{V_{n+1}^{(1)}}^{-1/2} x^{(1)}(t)\right)^2\right)^{1/2} \\
&\leq& \norm{v_1}{2} \norm{v_2}{2} \left(1-2\rho_0\right) ,\label{eqnarray2}
\end{eqnarray}
because according to Lemma \ref{stablemin}, \eqref{consistencycondition2} implies
\begin{eqnarray*}
	&& \sum\limits_{t=m+1}^{n} \left(v_1'{V_{n+1}^{(1)}}^{-1/2} x^{(1)}(t)\right)^2 \\
	&=& v_1'{V_{n+1}^{(1)}}^{-1/2} \left( {V_{n+1}^{(1)}} - {V_{m+1}^{(1)}} \right){V_{n+1}^{(1)}}^{-1/2} v_1 \\
	&\leq& \norm{v_1}{2}^2 \left(1- \eigmin{{V_{n+1}^{(1)}}^{-1/2} {V_{m+1}^{(1)}} {V_{n+1}^{(1)}}^{-1/2}}\right) \\
	&\leq& \norm{v_1}{2}^2 \left(1- \frac{\eigmin{{V_{m+1}^{(1)}} }}{\eigmax{{V_{n+1}^{(1)}}}}\right) \\
	&\leq& \norm{v_1}{2}^2 \left(1- \frac{m \eigmin{K_1}}{3n\eigmax{K_1}}\right) \\
	&\leq& \norm{v_1}{2}^2 \left(1- \frac{ \eigmin{K_1}}{9\eigmax{K_1}}\right) = \norm{v_1}{2}^2 \left(1-2 \rho_0\right)^2.
\end{eqnarray*}
Thus, by \eqref{eqnarray1} and \eqref{eqnarray2}, for an arbitrary unit vector $v$ we have
\begin{eqnarray*}
	v'E_nv \geq \norm{v_1}{2}^2 + \norm{v_2}{2}^2 - 2 \norm{v_1}{2} \norm{v_2}{2} \left(\rho_0 + 1 - 2 \rho_0 \right) \\
	= \rho_0 \left(\norm{v_1}{2}^2 + \norm{v_2}{2}^2\right) + \left(1-\rho_0 \right) \left(\norm{v_1}{2} - \norm{v_2}{2}\right)^2 ,
\end{eqnarray*} 
i.e. \eqref{consistencyeq1} holds. 
\endproof

\subsection{\bf Proof of Proposition \ref{consistencyaux2}}
Since
\begin{eqnarray*}
	V_{n+1}^{(2)} - \Sigma_n = \sum\limits_{t=m+1}^{n} \trans{2}^{t} \left(z(t)z(t)' - z(m)z(m)'\right) {\trans{2}'}^{t} ,
\end{eqnarray*}
and for $m+1 \leq t \leq n$, according to \eqref{Jordaneq1},
\begin{eqnarray*}
	&& \norm{z(t)-z(m)}{2} \\
	&\leq& \sum\limits_{i=m+1}^{t} \norm{\trans{2}^{-i}w^{(2)}(i)}{2} \\ 
	&\leq& \Opnorm{P^{-1}}{\infty}{2} \Mnorm{P}{\infty} \noisemax{n+1}{\delta} \sum\limits_{i=m+1}^{t} \MJordanconst{i}{\Lambda_2^{-1}} \\
	&\leq& \Opnorm{P^{-1}}{\infty}{2} \Mnorm{P}{\infty} \noisemax{n+1}{\delta} e^{\eigmin{\trans{2}}} \\
	&\times& t^{\mult{\trans{2}}} \eigmin{\trans{2}}^{-m-1},
\end{eqnarray*}
using Proposition \ref{zinfinitynormbound}, by \eqref{consistencycondition4} we have
\begin{eqnarray*} 
	&& \eigmax{\trans{2}^{-n} \left(V_{n+1}^{(2)} - \Sigma_n\right) {\trans{2}'}^{-n}} \\
	&\leq& 2\zinfinitybound{\trans{2}}{\delta} \sum\limits_{t=m+1}^{n} \norm{z(t)-z(m)}{2} \Mnorm{{\trans{2}'}^{-n+t}}{2}^2 \\
	&\leq& 2 \zinfinitybound{\trans{2}}{\delta}  \MJordanconst{}{{\trans{2}'}^{-1}}^2 \max\limits_{m+1 \leq t \leq n} \norm{z(t)-z(m)}{2} \\
	&\leq& \frac{1}{4} \rho_2 \minpoly{\trans{2}}^2\innerproductmin{\trans{2}}{\delta}^2 \noisemax{n+1}{\delta} n^{\mult{\trans{2}}} \eigmin{\trans{2}}^{-n/3} \\
	&\leq& \frac{1}{2}\eigmin{\trans{2}^{-n} V_{n+1}^{(2)} {\trans{2}'}^{-n}} .
\end{eqnarray*}

The last inequality above, is implied by \eqref{consistencycondition4}. Hence, 
\begin{eqnarray*} 
	\eigmax{V_{n+1}^{(2)} - \Sigma_n} \leq \frac{1}{2}\eigmin{V_{n+1}^{(2)}},
\end{eqnarray*}
which implies 
\begin{eqnarray*}
	&& \Mnorm{\Sigma_n^{1/2} {V_{n+1}^{(2)}}^{-1/2}}{2}^2 \\
	&=& \eigmax{I_{p_2}+ {V_{n+1}^{(2)}}^{-1/2} \left(\Sigma_n - {V_{n+1}^{(2)}}\right) {V_{n+1}^{(2)}}^{-1/2}} \\
	&\leq& \frac{3}{2}.
\end{eqnarray*}
Finally,
\begin{eqnarray*}
\Mnorm{\tilde{U}_n^{-1}U_n}{2}^2 \leq \Mnorm{n^{1/2}{V_{n+1}^{(1)}}^{-1/2}}{2}^2 + \Mnorm{\Sigma_n^{1/2} {V_{n+1}^{(2)}}^{-1/2}}{2}^2 \\
\leq \frac{2}{\eigmin{K_1}} + \frac{3}{2} .
\end{eqnarray*}
\endproof

\subsection{\bf Proof of Proposition \ref{consistencyaux3}}

For $t=0, \cdots, n+1$, define the sigma-fields $\mathcal{F}_t = \sigma \left(w(1),\cdots,w(t)\right)$. Letting $\symmetrizer{\cdot}$ be as defined in the proof of Lemma \ref{stablemin}, and $X_t=\symmetrizer{w(t+1) x^{(1)}(t)'}$ be a martingale difference sequence of symmetric matrices with respect to $\{\mathcal{F}_t\}_{t=0}^n$, all matrices 
\begin{eqnarray*}
	p \left(\MJordanconst{}{\trans{1}} \left(\norm{x^{(1)}(0)}{\infty}+ \noisemax{n+1}{\delta}\right) \noisemax{n+1}{\delta}\right)^2 I_{p+p_1} - X_t^2
\end{eqnarray*} 
are by Proposition \ref{statenorm} positive semidefinite. Letting 
\begin{eqnarray*}
	\sigma^2=p \MJordanconst{}{\trans{1}}^2 \left(\norm{x^{(1)}(0)}{\infty}+ \noisemax{n+1}{\delta}\right)^2 \noisemax{n+1}{\delta}^2 \left(n+1\right), 
\end{eqnarray*}
according to Proposition \ref{MAzuma}, by \eqref{consistencycondition5} we get \eqref{consistencyeq4} since
\begin{eqnarray*} 
\PP{\Mnorm{G_n}{2} > \frac{\epsilon}{\rho_3}} = \PP{\eigmax{\sum\limits_{t=0}^{n}X_t} > n\frac{\epsilon}{\rho_3}} \\
\leq 2 \left(p+p_1\right) \exp \left(-\frac{n^2 \epsilon^2}{8 \sigma^2 \rho_3^2}\right) \leq \frac{\delta}{2}.
\end{eqnarray*}
Next, for $\Mnorm{H_n}{2}$, using \eqref{Jordaneq1} and Proposition \ref{zinfinitynormbound},
\begin{eqnarray*}
	&& \Mnorm{n^{-1/2} \sum\limits_{t=0}^{m-1} w(t+1) x^{(2)}(t)' \Sigma_n^{-1/2}}{2} \\
	&\leq& n^{-1/2} \sum\limits_{t=0}^{m-1} \norm{w(t+1)}{2} \norm{z(t)}{2} \Mnorm{\Sigma_n^{-1/2}\trans{2}^t}{2} \\
	&\leq& p^{1/2}n^{-1/2} \noisemax{n+1}{\delta} \zinfinitybound{\trans{2}}{\delta} \\ 
	&\times& \Mnorm{\Sigma_n^{-1/2}\trans{2}^n}{2} \sum\limits_{t=n-m+1}^{n} \Mnorm{\trans{2}^{-t}}{2} \\
	&\leq& \rho_4 \noisemax{n+1}{\delta} n^{\mult{\trans{2}}-1/2} \eigmin{\trans{2}}^{-2n/3}.
\end{eqnarray*}
Thus, by \eqref{consistencycondition6} we have
\begin{eqnarray} \label{consistencyeq5}
\Mnorm{n^{-1/2} \sum\limits_{t=0}^{m-1} w(t+1) x^{(2)}(t)' \Sigma_n^{-1/2}}{2} \leq \frac{\epsilon}{3 \rho_3}.
\end{eqnarray}
Moreover, for $t=m,\cdots,n$, letting 
\begin{equation*}
\tilde{X}_t = \symmetrizer{w(t+1) x^{(2)}(m)' {\trans{2}'}^{t-m} \Sigma_n^{-1/2}}
\end{equation*}
be a martingale difference sequence with respect to $\{\mathcal{F}_t\}_{t=m}^n$ (note that both $\Sigma_n$ and $x^{(2)}(m)$ are $\mathcal{F}_m$ measurable), all matrices 
\begin{eqnarray*}
	\left( p^{1/2} \noisemax{n+1}{\delta} \norm{\Sigma_n^{-1/2}\trans{2}^{t-m}x^{(2)}(m)}{2} \right)^2 I_{p+p_2} - \tilde{X}_t^2
\end{eqnarray*} 
are positive semidefinite, so, according to Proposition \ref{MAzuma}, we have
\begin{eqnarray*}
	\PP{\eigmax{\sum\limits_{t=m}^{n} \tilde{X}_t} > \frac{n^{1/2}\epsilon}{3\rho_3} \Biggl| \mathcal{F}_m} \\
	\leq 2 \left(p+p_2\right) \exp \left(- \frac{n \epsilon^2}{72 \sigma^2 \rho_3^2}\right),
\end{eqnarray*}
where
\begin{eqnarray*}
	\sigma^2 &=& \sum\limits_{t=m}^{n} \left( p^{1/2} \noisemax{n+1}{\delta} \norm{\Sigma_n^{-1/2}\trans{2}^{t-m}x^{(2)}(m)}{2} \right)^2 \\
	&=& p \noisemax{n+1}{\delta}^2 \sum\limits_{t=m}^{n} \left(\trans{2}^{t-m}x^{(2)}(m)\right)'\Sigma_n^{-1}\trans{2}^{t-m}x^{(2)}(m) \\
	&=& p \noisemax{n+1}{\delta}^2 \tr{\Sigma_n^{-1} \sum\limits_{t=m}^{n} \trans{2}^{t-m}x^{(2)}(m) x^{(2)}(m)' {\trans{2}'}^{t-m}} \\
	&\leq&  p^2 \noisemax{n+1}{\delta}^2.
\end{eqnarray*}
The last inequality above is simply implied by the definition of $\Sigma_n$. Next, applying \eqref{consistencycondition7}, with probability at least $1-\delta/2$ it holds that 
\begin{eqnarray} \label{consistencyeq7}
\Mnorm{\sum\limits_{t=m}^{n} w(t+1) x^{(2)}(m)' {\trans{2}'}^{t-m} \Sigma_n^{-1/2}}{2} > \frac{n^{1/2}\epsilon}{3\rho_3}.
\end{eqnarray}
Since for $t=m, \cdots, n$,
\begin{eqnarray*}
	&& \norm{\Sigma_n^{-1/2}x^{(2)}(t) - \Sigma_n^{-1/2} \trans{2}^{t-m}x^{(2)}(m)}{2} \\
	&=& \norm{\Sigma_n^{-1/2}\trans{2}^{n} \trans{2}^{-n+t}\left(z(t) - z(m)\right)}{2} \\
	&\leq& \eigmin{\trans{2}^{-n}\Sigma_n {\trans{2}'}^{-n}}^{-1/2} \norm{ \sum\limits_{i=m+1}^{t} \trans{2}^{-n+t-i}w^{(2)}(i)}{2} \\
	&\leq& \frac{ \Opnorm{P^{-1}}{\infty}{2} \Mnorm{P}{\infty} e^{\eigmin{\trans{2}}} \noisemax{n+1}{\delta} n^{\mult{\trans{2}}} \eigmin{\trans{2}}^{-m-1}}{\eigmin{\trans{2}^{-n}\Sigma_n {\trans{2}'}^{-n}}^{1/2}} \\
	&\leq& \rho_5 \noisemax{n+1}{\delta} n^{\mult{\trans{2}}} \eigmin{\trans{2}}^{-n/3},
\end{eqnarray*}
by \eqref{consistencycondition8}, 
\begin{eqnarray*}
	\Mnorm{ \sum\limits_{t=m}^{n} w(t+1) \left(x^{(2)}(t)' - x^{(2)}(m)' {\trans{2}'}^{t-m} \right)  \Sigma_n^{-1/2} }{2} \leq \frac{n^{1/2}\epsilon}{3\rho_3}.
\end{eqnarray*}
So, \eqref{consistencyeq7} implies that the following holds, with probability at least $1-\frac{\delta}{2}$.
\begin{eqnarray*}
	\Mnorm{n^{-1/2} \sum\limits_{t=m}^{n} w(t+1) x^{(2)}(t)' \Sigma_n^{-1/2}}{2} \leq \frac{2\epsilon}{3\rho_3},
\end{eqnarray*}
which, in addition to \eqref{consistencyeq5}, yields \eqref{consistencyeq8}.
\endproof

\end{document}